\documentclass[backref, breaklinks, twocolumn, colorlinks]{aastex63}
\hypersetup{linkcolor=blue,citecolor=blue,filecolor=cyan,urlcolor=magenta}
\usepackage{enumitem}
\usepackage{booktabs}
\usepackage[fleqn]{amsmath}
\usepackage{amssymb}
\usepackage{tabularx}

\received{2022 March 24} \revised{2022 June 9} \accepted{2022 June 13}
\submitjournal{ApJ}

\shorttitle{The Fundamental Plane Relation for Gamma-Ray Pulsars Implied by 4FGL}
\shortauthors{Kalapotharakos et al.} 

\newcommand{\g}{\gamma}

\newcommand{\ec}{\epsilon_{\rm cut}}

\newcommand{\ed}{\dot{\mathcal{E}}}

\begin{document}
\tighten
\title{The Fundamental Plane Relation for Gamma-Ray Pulsars Implied by 4FGL}

\correspondingauthor{Constantinos Kalapotharakos}
\email{constantinos.kalapotharakos@nasa.gov,ckalapotharakos@gmail.com}

\author[0000-0003-1080-5286]{Constantinos Kalapotharakos}
\affiliation{University of Maryland, College Park (UMCP/CRESST II)\\
College Park, MD 20742, USA} \affiliation{Astrophysics Science
Division, NASA/Goddard Space Flight Center\\ Greenbelt, MD 20771,
USA}

\author[0000-0002-9249-0515]{Zorawar Wadiasingh}
\affiliation{University of Maryland, College Park (UMCP/CRESST II)\\
College Park, MD 20742, USA} \affiliation{Astrophysics Science Division,
NASA/Goddard Space Flight Center\\ Greenbelt, MD 20771, USA}

\author[0000-0001-6119-859X]{Alice K. Harding}
\affiliation{Theoretical Division, Los Alamos National Laboratory\\ Los Alamos, NM 87545, USA}

\author[0000-0002-7435-7809]{Demosthenes Kazanas} \affiliation{Astrophysics Science
Division, NASA/Goddard Space
Flight Center\\ Greenbelt, MD 20771, USA} 




\begin{abstract}
We explore the validity of the recently reported fundamental plane (FP) relation of $\gamma$-ray pulsars using 190 pulsars included in the latest 4FGL-DR3 catalog. This sample number is more than twice as large as that of the original study. The FP relation incorporates 4 parameters, i.e., the spin-down power, $\ed$, the surface magnetic field, $B_{\star}$, the total $\gamma$-ray luminosity, $L_{\gamma}$, and a spectral cutoff energy, $\epsilon_{\rm cut}$. The derivation of $\epsilon_{\rm cut}$ is the most intriguing one because $\epsilon_{\rm cut}$ depends on the proper interpretation of the available phase-averaged spectra. We construct synthetic phase-averaged spectra, guided by the few existing phase-resolved ones, to find that the best fit cutoff energy, $\epsilon_{\rm c1}$, corresponding to a purely exponential cutoff (plus a power law) spectral form, is the parameter that optimally probes the maximum cutoff energy of the emission that originates from the core of the dissipative region, i.e., the equatorial current sheet. Computing this parameter for the 190 4FGL pulsars, we find that the resulting FP relation, i.e. the $\gamma$-ray luminosity in terms of the other observables, reads $L_{\gamma}=10^{14.3\pm 1.3}(\epsilon_{\rm c1}/{\rm MeV})^{1.39\pm0.17}(B_{\star}/{\rm G})^{0.12\pm 0.03}(\ed/{\rm erg\;s^{-1}})^{0.39\pm 0.05}{\rm ~erg\;s^{-1}}$; this is in good agreement with both the empirical relation reported by \citet{2019ApJ...883L...4K} and the theoretically predicted relation for curvature radiation. Finally, we revisit the radiation reaction limited condition, to find it is a sufficient but not necessary condition for the theoretical derivation of the FP relation. However, the assumption of the radiation reaction limited acceleration reveals the underlying accelerating electric field component and its scaling with $\ed$.
\end{abstract}



\defcitealias{2013ApJS..208...17A}{2PC}
\defcitealias{2020ApJS..247...33A}{4FGL}
\defcitealias{2022arXiv220111184F}{4FGL-DR3}
\defcitealias{2005AJ....129.1993M}{ATNF}
\defcitealias{2019ApJ...883L...4K}{K19}

\section{Introduction} \label{sec:intro}

The Fermi Gamma-Ray Space Telescope has provided an unprecedented level of information, deepening our understanding of the $\gamma$-ray pulsars considerably. The observational data
compiled in the Second Fermi Pulsar Catalog \citep[2PC,
][]{2013ApJS..208...17A} showed various correlations especially
between spectral properties {and observed or derived parameters, such as cutoff energy, $\ec$, with light cylinder (LC) magnetic field, $B_{\rm LC}$ and spectral photon index, $\Gamma$, with spin-down power, $\ed$}, that actually pointed out the operational regime of {particle acceleration and $\gamma$-ray emission in pulsar magnetospheres} \citep{2017ApJ...842...80K}.

Recently, \citet[][hereafter \citetalias{2019ApJ...883L...4K}]{2019ApJ...883L...4K} demonstrated that the entire
$\gamma$-ray pulsar population (i.e., millisecond pulsars, MP and
young pulsars, YP) {obeys a luminosity function,}
%
\begin{equation}
\label{eq:fp_observations} L_{\gamma}\propto \ec^{1.18\pm
0.24}B_{\star}^{0.17\pm 0.05}\ed^{0.41\pm 0.08}
\end{equation}
that relates their total $\gamma$-ray luminosity, $L_{\gamma}$,
spectral cutoff energy, $\ec$, stellar surface magnetic field,
$B_{\star}$, and $\ed$. This constitutes a
pulsar ``fundamental plane" (FP) in these observationally derivable quantities -- $B_{\star} \propto \sqrt{P \dot{P}}$, $\ed \propto \dot{P}/P^3$, {where $P$ and $\dot{P}$ are the period and its first time derivative}, and $\ec$ the observed characteristic energy scale of the highest energy photons (see \S2 below). Remarkably, the observed FP
is close to the theoretical relation
\begin{equation}
\label{eq:fp_theory} L_{\gamma}\propto
\ec^{4/3}B_{\star}^{1/6}\ed^{5/12}
\end{equation}
{corresponding to curvature radiation (CR) emission in the radiation reaction limited (RRL) regime that occurs at the equatorial current sheet in the vicinity of the LC.} The FP was independently confirmed by \citet{2020JCAP...12..035P} who studied MPs, exploring the possibility that
unresolved MPs are responsible for the observed {GeV photon G}alactic center excess
{(\citealt{Abazajian2011}; also see, \citealt{2015ApJ...812...15B,2016PhRvL.116e1102B,Hooper_2016,2017ApJ...840...43A,2018ApJ...863..199G,2021arXiv210908439P,2021PhRvD.104d3007B,2021arXiv210600222G}}).

In \citetalias{2019ApJ...883L...4K}, {we} derived the observed FP, i.e., Eq.~\eqref{eq:fp_observations}, relying on the
\citetalias{2013ApJS..208...17A} data. More specifically, we used 88 pulsars
with reliable (i.e., published) spectral values (i.e., $L_{\gamma}, \ec$) out
of the 114 pulsars that were totally compiled in
\citetalias{2013ApJS..208...17A} while currently, more than 270 $\gamma$-ray
pulsars have been detected\footnote{https://confluence.slac.stanford.edu/display/GLAMCOG/\\Public+List+of+LAT-Detected+Gamma-Ray+Pulsars}.

On the one hand, the exponents in the expressions in Eqs.~(\ref{eq:fp_observations})-(\ref{eq:fp_theory}) indicate greater sensitivity of $L_{\gamma}$ on $\ec$ than $B_{\star}$ and $\ed$. On the other hand, the $B_{\star}$ and $\ed$ values of Fermi pulsars cover a range spanning several orders of magnitude while the corresponding $\ec$ values cover a limited range {of less than an order of magnitude}. Thus, the determination of $\ec$ becomes vital for a reliable derivation of the FP. The \citetalias{2013ApJS..208...17A} $\ec$ values, employed in \citetalias{2019ApJ...883L...4K}, had been derived adopting a power law with a pure exponential cutoff spectral function model. Nonetheless, the Fermi fits indicate that for the brightest pulsars, a sub-exponential function model can statistically describe better the phase-averaged spectra.

In this paper, we first explore the behavior of these different function models, uncovering the parameter values that best correlate with the cutoff energies corresponding to the highest-energy pulsar emission. Then, we update the FP using the spectral analysis data from the 12-year incremental version of the fourth Fermi-LAT catalog of point $\gamma$-ray sources (4FGL, \citealt{2020ApJS..247...33A}; 4FGL-DR3, for Data Release 3, \citealt{2022arXiv220111184F}) combined with data from the Australia Telescope National Facility (ATNF) Pulsar Catalog
\citep{2005AJ....129.1993M}\footnote{Updated catalog data can be found in
\url{https://www.atnf.csiro.au/research/pulsar/psrcat/}} that expand by more than two fold (compared to our previous exploration) the sample of pulsars. Finally, we revisit and further generalize the necessary and sufficient underlying theoretical conditions put forth in \citetalias{2019ApJ...883L...4K}.

The structure of the paper is as follows. In Section~\ref{sec:spectrum_function_model}, we discuss the behavior of the spectral fitting function models and the related parameters. In Section~\ref{sec:fundamental plane_observations}, we present the expanded sample of Fermi pulsars taking into account the \citetalias{2022arXiv220111184F} and \citetalias{2005AJ....129.1993M} data and update the FP relation of $\gamma$-ray pulsars. In section~\ref{sec:FPrevisited}, we review the RRL requirement and discuss the possible effects on the FP. Finally, in the last section, we discuss our results and summarize our conclusions.

\section{Finding the cutoff energy} \label{sec:spectrum_function_model}

Ideally, a fitting function model should describe the spectra and provide meaningful parameter values that correspond to the underlying physics. A model that describes the phase-averaged energy-weighted differential spectral energy density (SED) adequately for the vast majority of Fermi pulsars reads
\begin{equation}
    \label{eq:spectrumfit_bfree}
    \epsilon^2\frac{dN}{d\epsilon}=A \epsilon^{2-\Gamma}\exp\left[-\left(\frac{\epsilon}{\epsilon_{\rm
    cut}}\right)^b\right]~.
    \vspace{0.0in}
\end{equation}
where $\epsilon$ is the photon energy, $dN$ is the number of photons in energy bin $(\epsilon, \epsilon+d\epsilon)$, and $\Gamma$ is the photon index. The parameter $b$ describes the decay of the emitted power at large energies. The pure exponential cutoff corresponds to $b=1$ while the brightest Fermi pulsars (e.g., Vela, Crab, Geminga) are well-described by a sub-exponential cutoff (i.e., $b<1$).

The apex energy, $\epsilon_{\rm A}$, that signifies the photon energy  of the SED maximum occurs at
\begin{equation}
\label{eq:apex_energy}
\epsilon_{\rm
A}=\left(\frac{2 - \Gamma}{b}\right)^{1/b}\ec~.
\end{equation}
Thus, for fixed $\Gamma<2$ and $b$ values, $\epsilon_{\rm A}$ traces $\ec$ (being always proportional to it). However, the Fermi data have shown that $\Gamma$ positively correlates with the spin-down power $\ed$ \citepalias{2013ApJS..208...17A}. This implies that following the behavior of $\ec$ with $\epsilon_{\rm A}$ no longer reveals the corresponding $\ec$ behavior. Therefore, $\ec$ can remain constant with $\ed$ and still $\epsilon_{\rm A}$ can decrease with $\ed$ (as long as $\Gamma$ increases with $\ed$).

\begin{figure}[!tbh]
\vspace{0.0in}
  \begin{center}
    \includegraphics[width=1.0\linewidth]{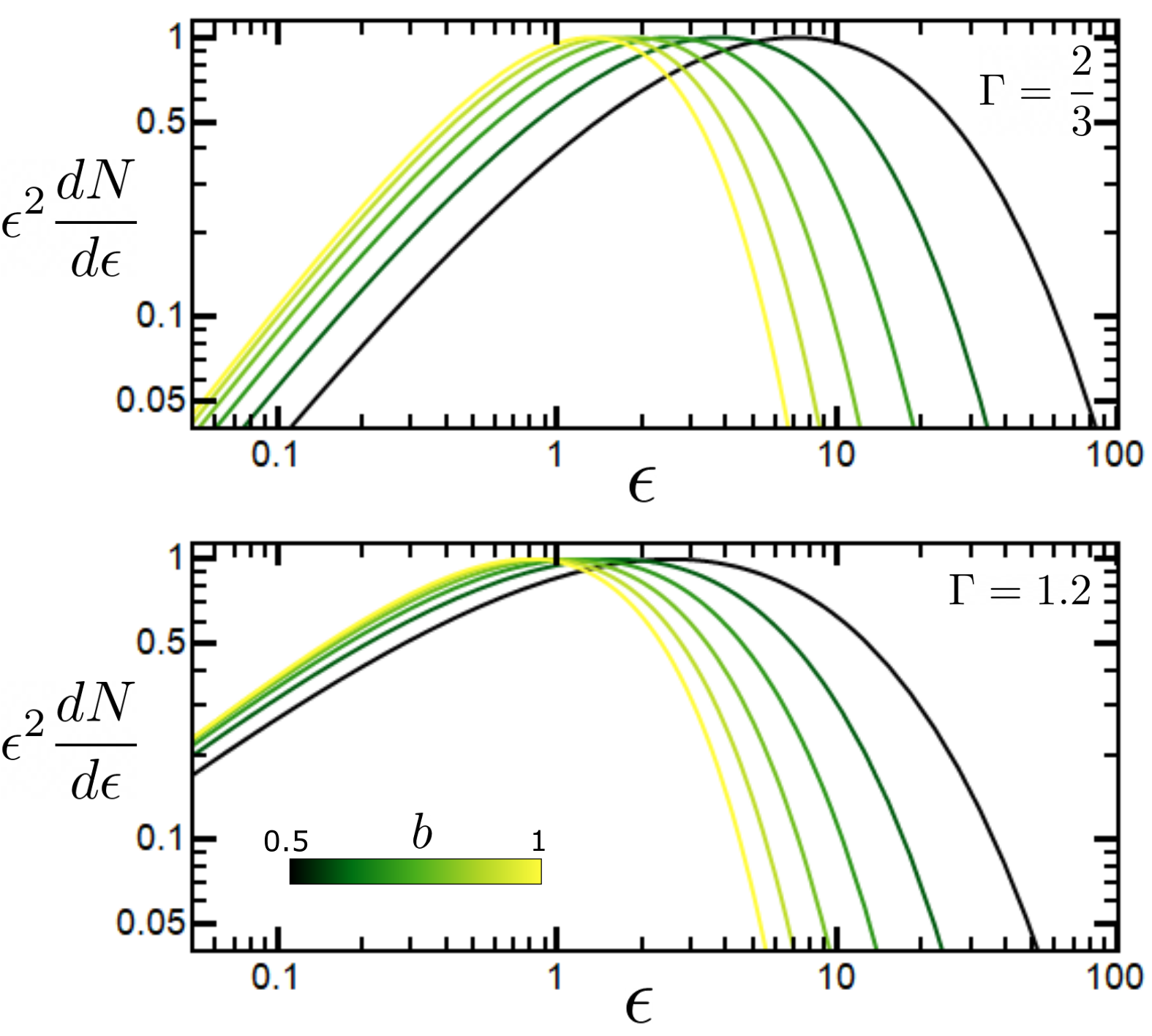}
  \end{center}
  \vspace{-0.2in}
  \caption{SED spectra corresponding to the function model in Eq.~\eqref{eq:spectrumfit_bfree} are plotted. The $\epsilon_{\rm cut}$ is fixed at 1 (in arbitrary units) while $b$ varies according to the indicated color scale. The two panels correspond to the indicated  $\Gamma$, i.e., photon index values. As $b$ decreases $\epsilon_{\rm cut}$ stops probing the apex energy becoming irrelevant to the energy level where the rapid decrease of the SED begins. The smaller the $\Gamma$ is the more prominent this effect is.}
  \label{fig:ecutirrel}
  \vspace{0.0in}
\end{figure}

Moreover, for $b\sim 1$, $\ec$ signifies the energy {at which the} rapid decline of the SED begins. However, when $b$ starts deviating from 1, the corresponding $\ec$ value {no longer discloses} that {characteristic photon energy scale}. In Figure~\ref{fig:ecutirrel}, we plot normalized ---to the maximum value--- mock SEDs color-coded according to $b$ values that range from 0.5 to 1 and for $\Gamma=2/3$ (top panel) and $\Gamma=1.2$ (bottom panel). As $b$ becomes gradually smaller than 1, the $\ec$ values, which in Fig.~\ref{fig:ecutirrel} are fixed at 1 (in arbitrary units), do not correspond to the apex energy and the energy level where the rapid decline of the curve occurs. The smaller the $\Gamma$ value, the more prominent this effect is (see top panel). Note that in \citetalias{2020ApJS..247...33A}, \citetalias{2022arXiv220111184F} even though function models with spectral shapes equivalent to Eq.~\eqref{eq:spectrumfit_bfree} are used, no characteristic energy level is mentioned, similarly to $\ec$, except the pivot energy in which the observed spectrum has the smallest error (see equation 4 in \citetalias{2020ApJS..247...33A} and Eq.~\ref{eq:4FGLspectrumfit} below).

\begin{figure}[!tbh]
\vspace{0.0in}
  \begin{center}
    \includegraphics[width=1.0\linewidth]{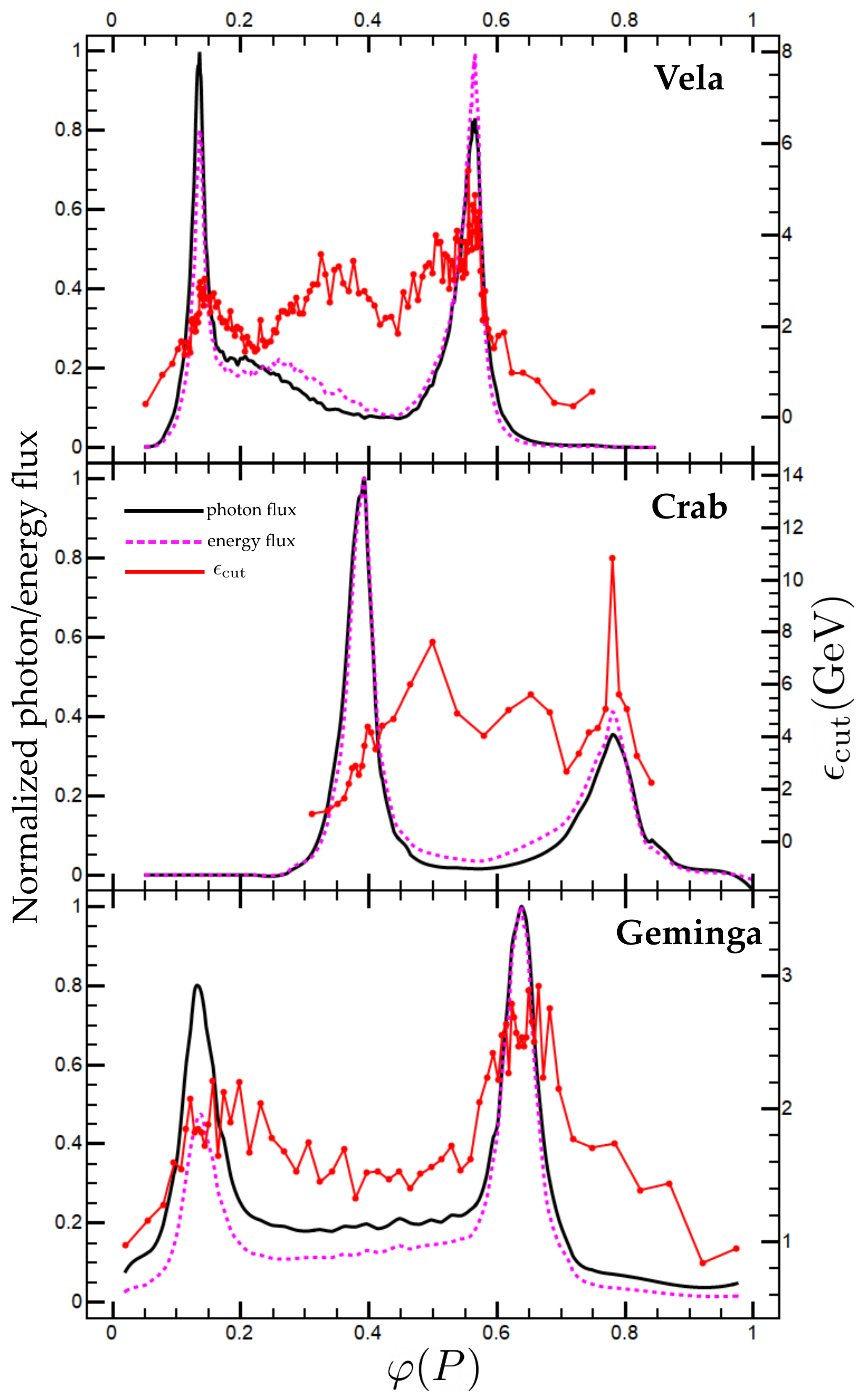}
  \end{center}
  \vspace{-0.2in}
  \caption{The Fermi $\gamma$-ray light curves together with the exponential $\epsilon_{\rm cut}$ energies of the corresponding phase-resolved spectra for the Vela, Crab, and Geminga pulsars as indicated in the figure. The black solid and magenta dashed lines denote the photon and energy flux, respectively (left-hand side ordinate). The red lines denote the exponential $\epsilon_{\rm cut}$ energies (right ordinate). The data have been adapted from \citet{2013PhDT.......182D}.}
  \label{fig:big3}
  \vspace{0.0in}
\end{figure}

\citet{2018ApJ...869L..18H},
\citet{2018NatAs...2..247T}, \citet{2019MNRAS.489.5494T}, and \cite{2021ApJ...923..194H} have shown that the
low-energy part of the Fermi spectrum is mainly {synchrocurvature radiation in the synchrotron radiation regime from
lower-energy particles while the higher-energy part of the spectrum is
synchrocurvature in the CR regime} of the high-energy particles. The  {theoretical scaling of Eq.~\eqref{eq:fp_theory}} assumes emission due to CR and thus,
the adopted $\ec$ in Eqs.~\eqref{eq:fp_observations},~\eqref{eq:fp_theory} should probe the characteristic {photon energy scale} of the high-energy part of the spectrum.

In that respect, the $\Gamma$ values and their dependence on $\ed$ reflect the lower-energy {particles'} properties (i.e., number and energy distribution). On the other hand, the LAT Vela paper, \citep{2010ApJ...713..154A}, showed that the observed sub-exponential cutoff of the phase-averaged spectra possibly reflects the superposition of many purely exponential cutoff spectra corresponding to different phases and magnetosphere regions.

In the phase-averaged spectra, the different phases involve photons from different magnetospheric locations corresponding, in general, to different electric fields, particles energies, and radii of curvature, which introduces ambiguity regarding the best choice for $\ec$ that should be considered for the $\gamma$-ray luminosity function, i.e., FP, relation.

\begin{figure}[!tbh]
\vspace{0.0in}
  \begin{center}
    \includegraphics[width=1.0\linewidth]{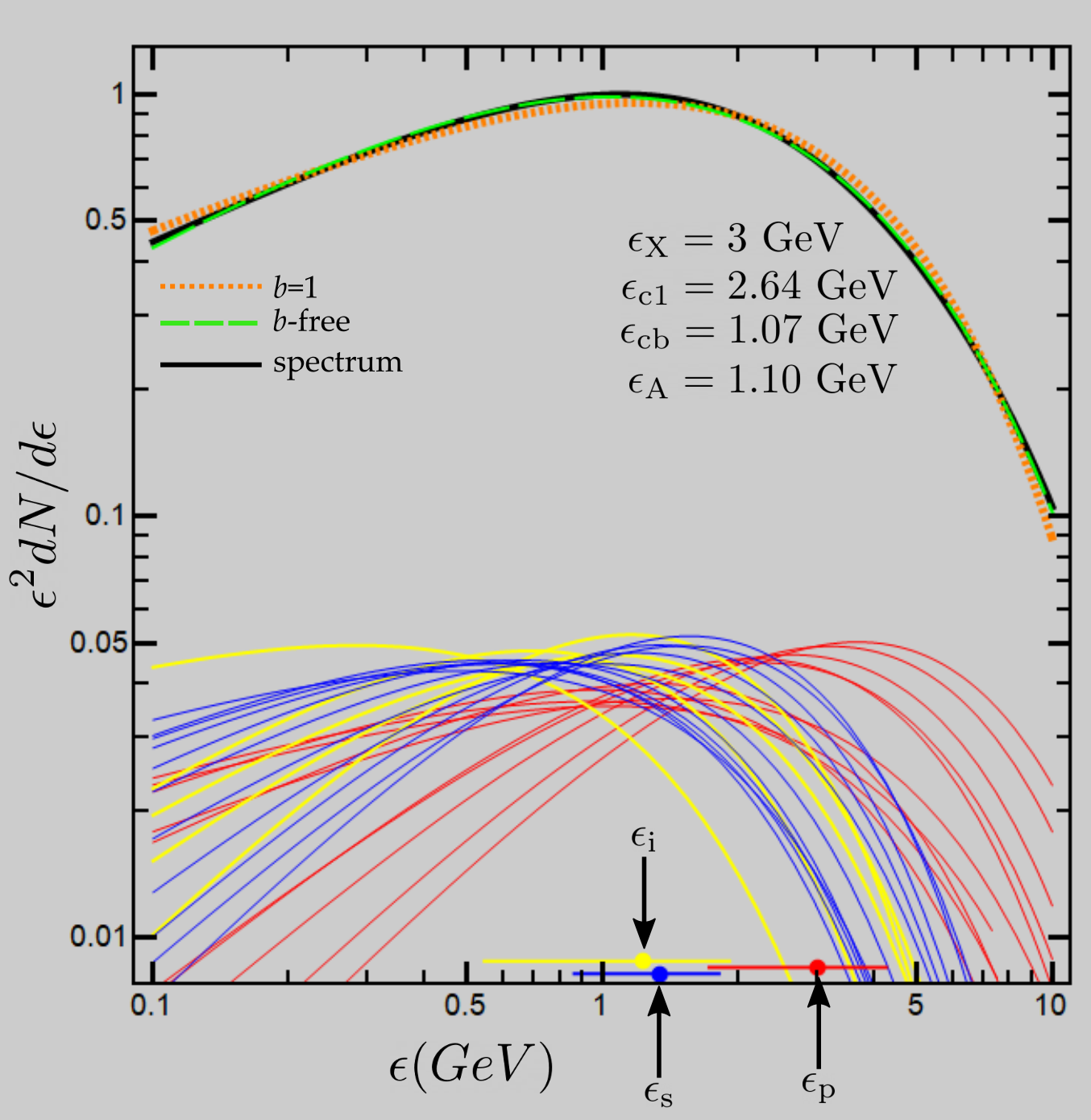}
  \end{center}
  \vspace{-0.2in}
  \caption{A phase-averaged synthetic spectrum (black line) and the corresponding fit lines considering $b$-free (green dashed line) and $b=1$ (orange dotted line). The red, blue, and yellow lines denote the corresponding phase-resolved contributing spectra of the primary, secondary, and inter/off peak emission components. The big dots and the horizontal lines near the bottom abscissa indicate the corresponding characteristic cutoff energies and ranges (see text for more details). For clarity only 1/4 of the contributing spectra has been plotted.}
  \label{fig:synthetic_spectrum_example}
  \vspace{0.0in}
\end{figure}

Phase-resolved spectral analysis for a number of bright pulsars (e.g., Vela, Crab, Geminga) have indicated that the $\ec$ values corresponding to the phases around the peaks of the $\gamma$-ray light curves tend to be higher \citep[][see Fig.~\ref{fig:big3}]{2013PhDT.......182D}\footnote{M. DeCesar's dissertation can be found in the digital repository at the University of Maryland \url{https://drum.lib.umd.edu/handle/1903/14595}.}. The peak phases probe the core-emitting magnetosphere regions considered to be at the so-called Y-point and in the equatorial current sheet emanating from it \citep{2010ApJ...715.1282B,2010MNRAS.404..767C,2014ApJ...793...97K,2016MNRAS.457.2401C,2018ApJ...855...94P,2018ApJ...857...44K}. In realistic oblique rotators, the axisymmetry breaks and therefore, the Y-point and equatorial current sheet regions are not equally efficient in producing $\gamma$-ray emission \citep{2014ApJ...793...97K,2016MNRAS.457.2401C,2018ApJ...855...94P,2018ApJ...857...44K}. Thus, for any specific object different observers trace, in principle, emission from different magnetosphere regions. Actually, in \citetalias{2019ApJ...883L...4K}, we claimed that the FP theory implies that many different (though parallel) FPs exist corresponding to these different magnetosphere regions, which, in general, incorporate different particle energies and radii of curvature. These planes contribute, partially to the observed scatter around the FP.

\begin{deluxetable*}{cc}
\label{tab:ecut_nomenclature}
\tablecaption{Cutoff energy nomenclature}


\tablehead{\colhead{Symbol} & \colhead{Description}}


\startdata
$\epsilon_{\rm cut}$ & Non-specific cutoff energy value \\
$\epsilon_{\rm c1}$ & Cutoff energy value corresponding to $b=1$ in Eq.~\eqref{eq:4FGLspectrumfit} \\
$\epsilon_{\rm cb}$ & Cutoff energy value corresponding to $b$ free in Eq.~\eqref{eq:4FGLspectrumfit} \\
$\epsilon_{\rm p}$ & Characteristic cutoff energy value of the primary peak emission \\ $\epsilon_{\rm s}$ & Characteristic cutoff energy value of the secondary peak emission \\ $\epsilon_{\rm i}$ & Characteristic cutoff energy value of the inter/off peak emission \\ $\epsilon_{\rm x}$ & $\max(\epsilon_{\rm p},\epsilon_{\rm s})$ \\
$\epsilon_{\rm A}$ & The SED apex energy \\
$\epsilon_{\rm f}$ & The energy where the SED power is $f$ times lower than at $\epsilon_{\rm A}$
\enddata
\end{deluxetable*}

Nonetheless, in typical {double-peaked pulse profiles}, the $\ec$ values corresponding to the different peaks are not always {indistinguishable}.  In phase-averaged spectra, the $\ec$ determination corresponding to any of the peaks is {nontrivial}. Below, we propose treatments of the phase-averaged spectrum that provide energy values that capture the highest exponential cutoff energies found around the Fermi $\gamma$-ray light-curve peaks, which supposedly trace the most efficient $\gamma$-ray emitting regions
for the observer angle values corresponding to the Earth's line of sight.

\subsection{Purely exponential cutoff energy}\label{sec:pure_exponential_cutoff}

\begin{figure}[!tbh]
\vspace{0.0in}
  \begin{center}
    \includegraphics[width=1.0\linewidth]{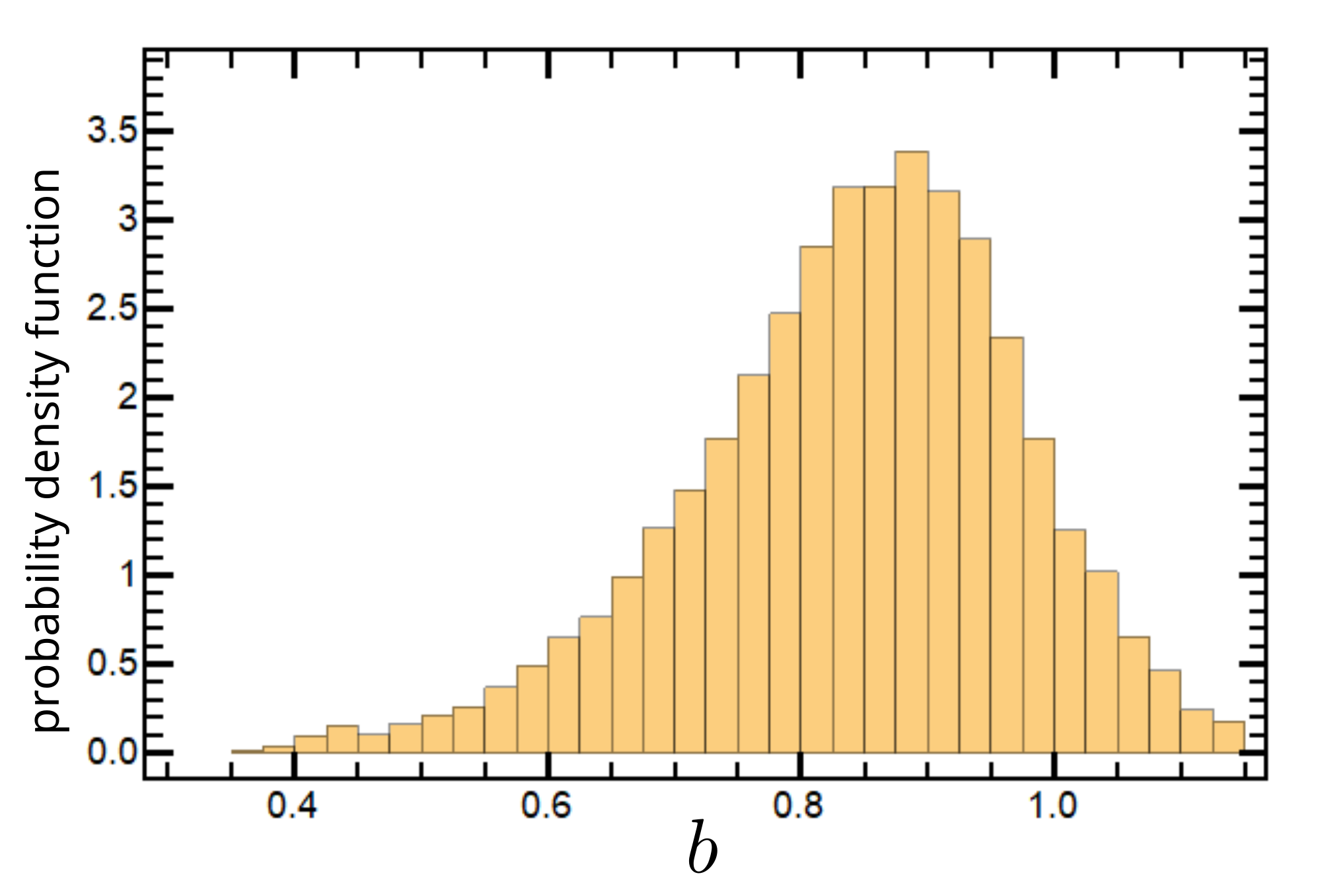}
  \end{center}
  \vspace{-0.2in}
  \caption{The probability density function histograms of $b$ for the $10^4$ synthetic spectra.}
  \label{fig:b_histogram}
  \vspace{0.0in}
\end{figure}

We explore here the characteristic energy levels that best probe the maximum $\epsilon_{\rm cut}$ value, at the peaks of the $\gamma$-ray light curves. Unfortunately, even though there are at least a few tens of Fermi pulsars that that are bright enough to allow phase-resolve spectroscopy, currently, there are only half a dozen of bright pulsars with published phase-resolved spectra \citep{2010ApJ...720..272A,2010ApJ...720...26A,2013PhDT.......182D}. Thus, in this study, we explore how various characteristic energies correlate with the maximum exponential cutoff energies in the $\gamma$-ray light-curve peaks using synthetic spectra.

To explore such behavior, we synthesize mock phase-resolved spectra making considerations that are guided by the phase-resolved spectroscopy results for bright pulsars \citep[][see also Figure~\ref{fig:big3}]{2013PhDT.......182D} which then determine the phase-averaged ones.

As mentioned above, the sub-exponential cutoffs can be the result of the superposition of many purely exponential spectra of varying purely exponential cutoff energies, $\epsilon_{\rm c1}$\footnote{For the rest of the paper the pure exponential cutoff energy of a spectrum corresponding to $b=1$ in the Eq.~\eqref{eq:spectrumfit_bfree} will be denoted by $\epsilon_{\rm c1}$ (see also Table~\ref{tab:ecut_nomenclature}).}, and $\Gamma$ values that correspond to different magnetosphere regions and particle populations. Thus, one of our simplified assumptions is that the phase-averaged spectra are the superpositions of many ($\sim 100$) purely exponential phase-resolved spectra, i.e., Eq.~\eqref{eq:spectrumfit_bfree} with $b=1$, that correspond to different phases.

The production of the simulated spectra does not depend so much on the details of the underlying shape of the $\gamma$-ray light curves and it only requires a proper (i.e., realistic) consideration of the various spectral components.
Thus, we consider that there are three major spectral components, which are those corresponding to two main peaks and one corresponding to the inter-peak and off-peak emission.

We generate $10^4$ phase-averaged
synthetic spectra in which the percentage of the power\footnote{The emitting power is the integrated energy of an emitting spectral component (e.g., primary peak) that is emitted within a stellar period over this period.} corresponding to the inter/off-peak emission (with respect to the total spectrum power) is randomly selected from a uniform distribution that ranges from 10 to 40\%. The percentages of the emitting power corresponding to the primary and secondary peaks for each synthetic spectrum are also randomly selected taking into account the constraints: \textbf{(i)} the primary peak carries at least half of the remaining power and \textbf{(ii)} the secondary peak carries a power that is never less than 20\% of the total power.

\begin{deluxetable*}{ccccccc}
\label{tab:big3_spectral_characteristics}
\tablecaption{Phase-resolved spectrum characteristics for the Vela, Crab, and Geminga pulsars}



\tablehead{
\colhead{} & \colhead{Primary peak} & \colhead{Secondary peak} & \colhead{Inter/off peak} &
\colhead{} &\colhead{} & \colhead{}\\
\colhead{Pulsar name} & \colhead{relative power} & \colhead{relative power} & \colhead{relative power} & \colhead{$\epsilon_{\rm p}$} & \colhead{$\epsilon_{\rm s}$} & \colhead{$\epsilon_{\rm i}$}\\
\colhead{} & \colhead{(\%)} & \colhead{(\%)}  & \colhead{(\%)} & \colhead{(GeV)} & \colhead{(GeV)} & \colhead{(GeV)}}


\startdata
 J0835-4510 (Vela) & 42 & 20 & 38 & \textbf{3.9} & 2.4 & 2.3 \\
 J0534+2200 (Crab) & 51 & 30 & 19 & 3.2 & \textbf{5.8} & 5.2 \\
 J0633+1746 (Geminga) & 44 & 22 & 34 & \textbf{2.6} & 1.9 & 1.6 \\
\enddata
\tablecomments{The boldface in $\epsilon_{\rm p},~\epsilon_{\rm s}$ denotes the highest cutoff energy, i.e., $\epsilon_{\rm X}$.}
\end{deluxetable*}

\begin{figure*}
\vspace{0.0in}
  \begin{center}
    \includegraphics[width=1.0\linewidth]{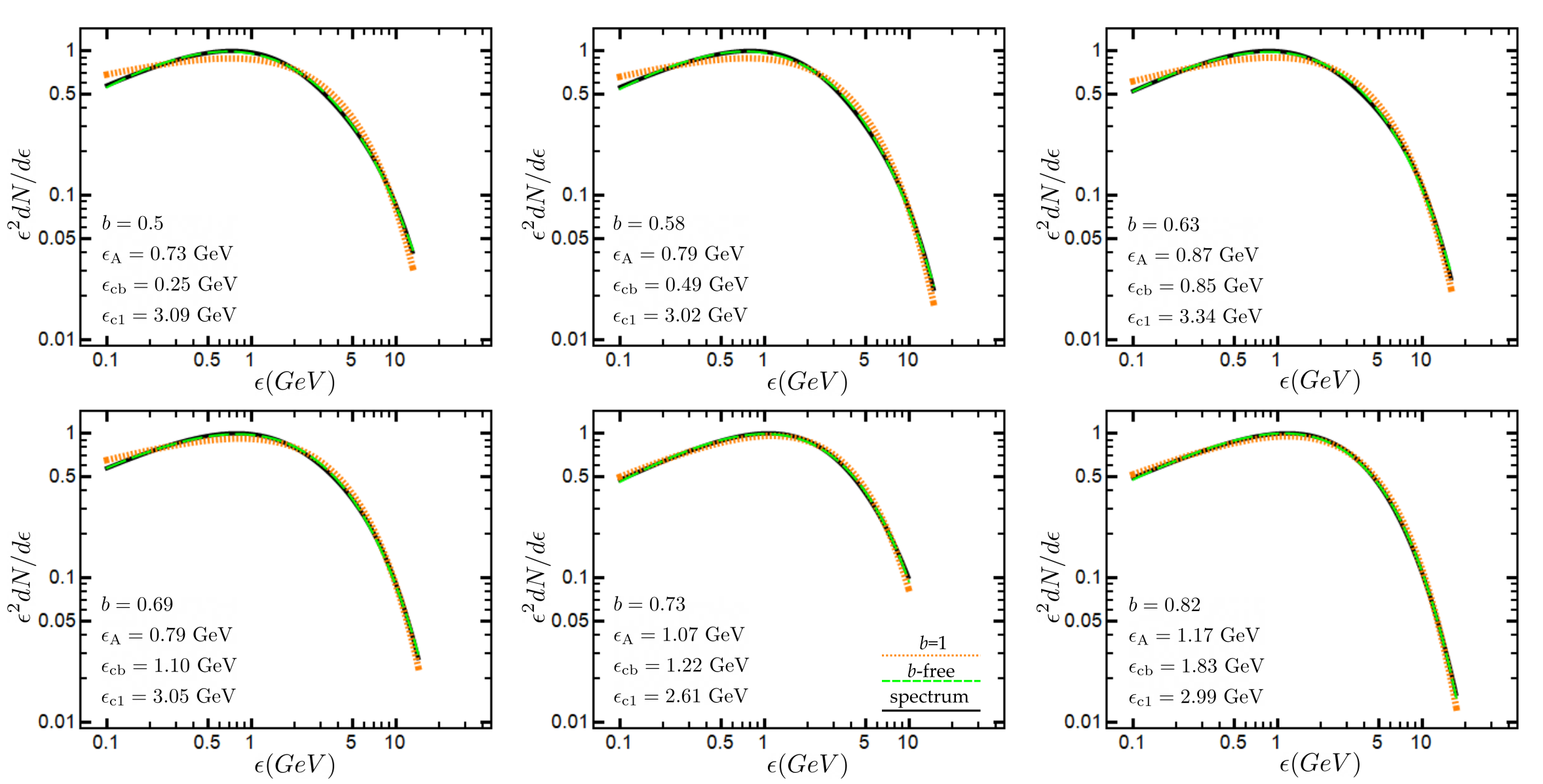}
  \end{center}
  \vspace{-0.2in}
  \caption{Typical examples of phase-averaged synthetic spectra (black lines), in which $\epsilon_{\rm X}=3$ GeV and the corresponding fits assuming the $b$-free function model (green dashed lines), and the fixed $b=1$ function model (orange dotted lines). The corresponding $b,~\epsilon_{\rm A},~\epsilon_{\rm cb}$, and $\epsilon_{\rm c1}$ are indicated in each panel. The $\epsilon_{\rm c1}$ values probe the $\epsilon_{\rm X}$ value very well.}
  \label{fig:spectra_examples}
  \vspace{0.0in}
\end{figure*}

We also assign characteristic (purely exponential) cutoff values, $\epsilon_{\rm p}$, $\epsilon_{\rm s}$, and $\epsilon_{\rm i}$ corresponding to the primary, secondary, and inter/off peak emission components, respectively. The maximum characteristic cutoff value among the two peaks $\epsilon_{\rm X}=\max(\epsilon_{\rm p},\epsilon_{\rm s})$ can be assigned to any of the two peaks (i.e., primary and secondary) and is selected randomly from a uniform distribution that ranges from 1 to 6 GeV. Note that the primary and secondary peaks are defined based on the corresponding power carried by each peak, i.e., the integrated energy under each peak over a period. We emphasize that this is not necessarily related to the peak amplitudes and the cutoff energies of the corresponding spectra. Observationally, for a majority of Fermi pulsars, the amplitude ratio P2/P1, where P1, P2 are the amplitude of the first and second peak (ordered by the phase lag from the radio peak), increases with the photon energy \citep[see][]{2010ApJ...720...26A}. This implies that the highest cutoff spectral energy appears in the second peak (see Figure~\ref{fig:big3}), which however does not guarantee that this peak, i.e., the second one, carries more energy. Thus, in the Crab pulsar, the first peak is the one that carries more energy than the second one despite the fact that the second peak has a higher spectral cutoff energy.

After the $\epsilon_{\rm X}$ value is selected and assigned to either $\epsilon_{\rm p}$ or $\epsilon_{\rm s}$, the characteristic value corresponding to the other peak is also randomly selected to be between 1/3 and 2/3 of $\epsilon_{\rm X}$, which ensures cases where $\epsilon_{\rm p}$ and $\epsilon_{\rm s}$ considerably differ and cases where they are close to each other. The phase-resolved spectra indicate the the cutoff energy values corresponding to the inter/off peak emission are either lower than the cutoff values of both peaks or in between these values. Thus, in our model, $\epsilon_{\rm i}$ is randomly selected to lie in the interval $[0.5\min(\epsilon_{\rm p},\epsilon_{\rm s}),4/5\epsilon_{\rm X}]$.

\begin{figure}[!tb]
\vspace{0.0in}
  \begin{center}
    \includegraphics[width=1.0\linewidth]{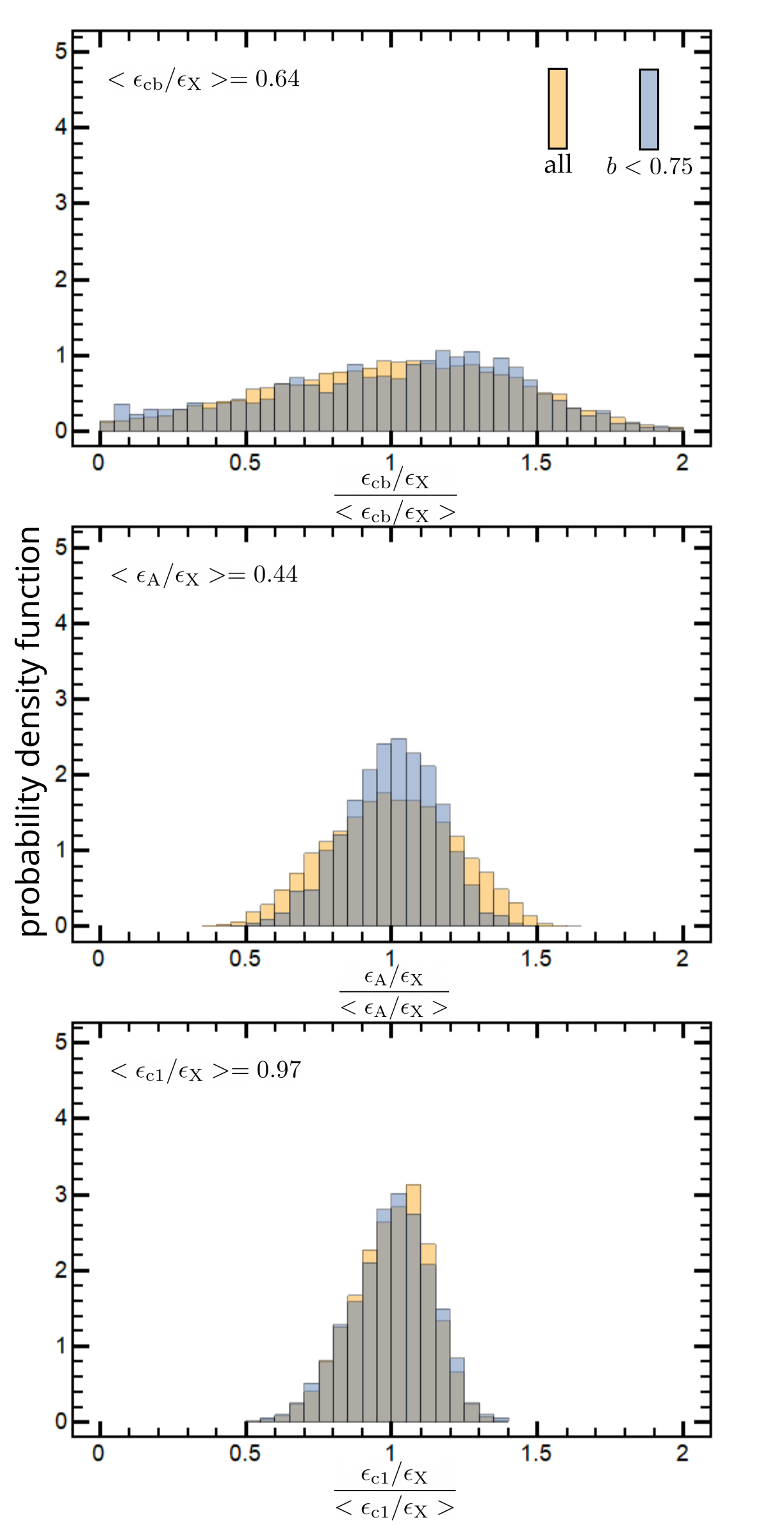}
  \end{center}
  \vspace{-0.2in}
  \caption{Probability density function histograms for the indicated energy ratios are shown. All the histograms are normalized with respect to the corresponding mean ratio value, which is indicated in each panel. The orange histogram refer to the entire population of the synthetic spectra while the blue ones refer to those with $b\leq 0.75$. The gray areas indicate overlapping.}
  \label{fig:ratio_histograms}
  \vspace{0.0in}
\end{figure}

\begin{figure}[!tb]
\vspace{0.0in}
  \begin{center}
    \includegraphics[width=1.0\linewidth]{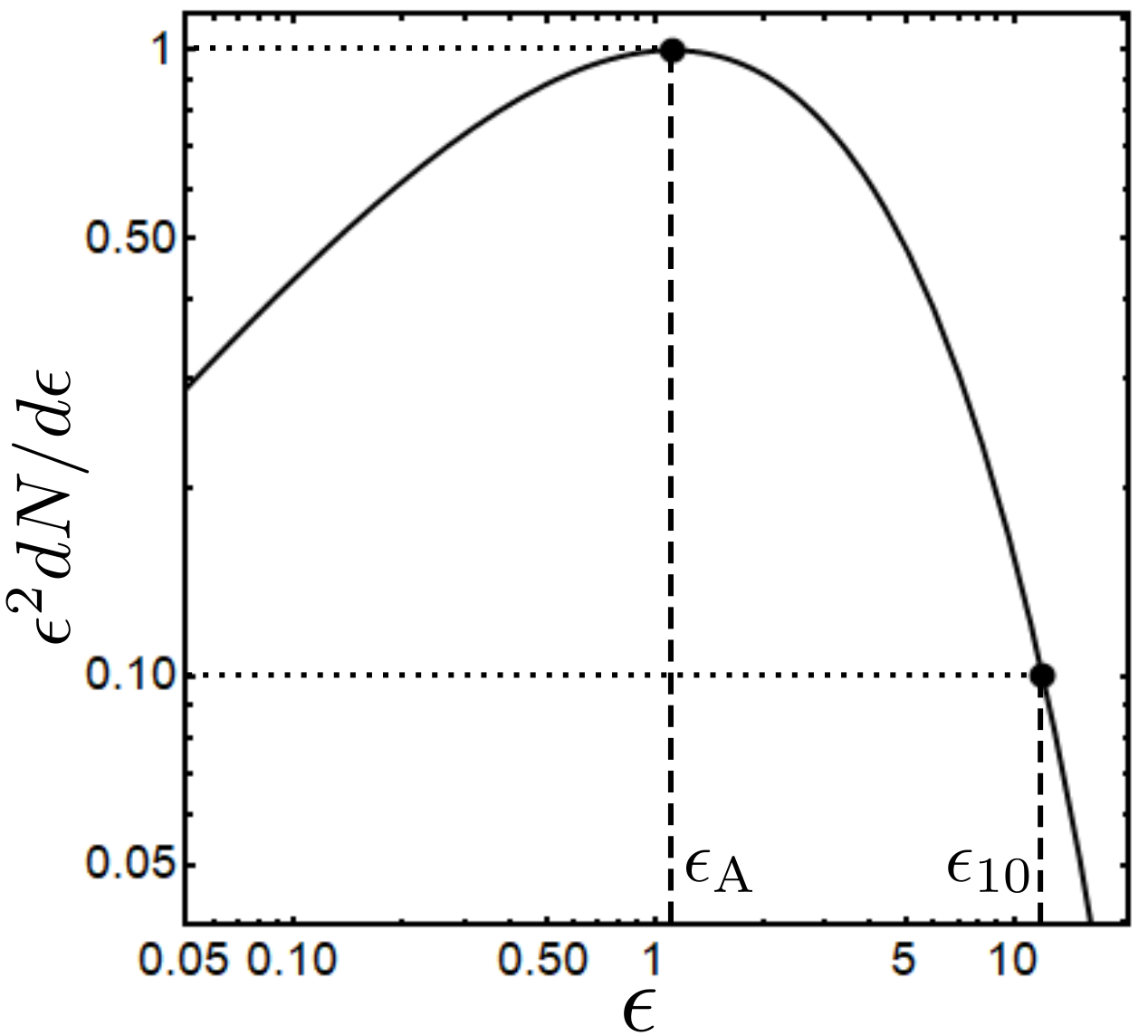}
  \end{center}
  \vspace{-0.2in}
  \caption{The $\epsilon_{10}$ value is the energy $\epsilon>\epsilon_{\rm A}$ where the SED power is 1/10 of the corresponding maximum power at $\epsilon_{\rm A}$.}
  \label{fig:en10_demo}
  \vspace{0.0in}
\end{figure}

Each of the major emission components runs over a phase interval consisting of spectra with a variety of cutoff energies, i.e., $\epsilon_{\rm c1}$ values, (e.g., Fig.~\ref{fig:big3}). Thus, for each emission component, we consider the contribution of, in general, several purely exponential spectra with $\epsilon_{\rm c1}$ values that are uniformly distributed over an $\epsilon$-interval around the corresponding characteristic values (i.e., $\epsilon_{\rm p}$, $\epsilon_{\rm s}$, $\epsilon_{\rm i}$). The length of these intervals are randomly selected to be between 1/3 and 2/3 of the corresponding characteristic values, emulating lower and greater variety of spectra within each emission component. We note that the $\Gamma$ values of the contributing purely exponential spectra are also randomly selected from the interval $[0.9,1.75]$, which is the interval the vast majority of Fermi pulsar lie.
We note that the adopted broad range of $\Gamma$ values is motivated by the broad range of $\Gamma$ values seen in the phase-resolved spectra \citep{2013PhDT.......182D}. In any case, it is noted that the results presented below regarding especially the conclusions of the cutoff energy parameters do not depend on the adopted $\Gamma$ values\footnote{We tested cases with different ranges and cases with the same $\Gamma$ values and attained similar results.}. We also highlight that monoenergetic $e^{\pm}$ produce spectra (from either curvature or synchrotron radiation) with a photon index $\Gamma=2/3$, which is quite different from what is observed in the available phase-resolved spectra \citep{2013PhDT.......182D}. The observed $\Gamma$ values in the phase-resolved spectra imply the existence of particle populations with a range of smaller energies, i.e., low-energy pairs, than the high-energy particles that are responsible for the high-energy part of the spectrum and the observed cutoff.

We note that for each component, the number of the purely exponential cutoff spectra, contributing to the synthetic spectrum, depends on the corresponding power percentage. So, for instance for an emission component whose emission power contributes 32\% to the total power, we consider 32 contributing (purely exponential cutoff) spectra of equal integrated energy. Superimposing all these spectra, corresponding to the 3 emission components, we construct the corresponding phase-averaged one. Then, we numerically identify the corresponding $\epsilon_{\rm A}$ value and fit the spectra using the function model in Eq.~\eqref{eq:spectrumfit_bfree} considering both the $b$ free\footnote{The cutoff energy value in Eq.~\eqref{eq:spectrumfit_bfree} corresponding to $b$-free fit will be denoted by $\epsilon_{\rm cb}$ in the rest of the paper (see also Table~\ref{tab:ecut_nomenclature}).} and $b=1$ cases. The fitting results also depend on the data extent in energy. Hence, in order to mimic the different data extents, all the spectra range from 0.1 GeV up to an energy that is selected randomly and is uniformly distributed from 8 to 30 GeV.

In our study, many cutoff energy values, denoted by different symbols, are considered. Thus, in Table~\ref{tab:ecut_nomenclature}, we present for clarity all the cutoff energy symbols used in this paper and their description.

In Table~\ref{tab:big3_spectral_characteristics}, we present the phase-resolved spectrum characteristic values for the Vela, Crab, and Geminga pulsars from data adapted from \citet{2013PhDT.......182D}. More specifically, we present the powers of the various emission components (i.e., primary, secondary, and inter/off peak) as percentages of the corresponding total powers and the corresponding $\epsilon_{\rm p}, \epsilon_{\rm s}$ and $\epsilon_{\rm i}$ values as these are derived by the phase-resolved spectroscopy (see Figure~\ref{fig:big3}). The properties of these three bright pulsars are commensurate with the choices made for generating synthetic spectra, discussed above.

In Figure~\ref{fig:synthetic_spectrum_example}, we demonstrate the case of one synthetic SED spectrum. The synthetic spectrum denoted by the black line is the result of the addition of 100 phase-resolved spectra. The contributing phase-resolved spectra corresponding to the primary peak, secondary peak, and the inter/off peak emission components are denoted by the red, blue, and yellow color lines, respectively. We note that, for clarity, we plot only one fourth of the contributing spectra, i.e., 25 phase-resolved spectra. The $\epsilon_{\rm p}$, $\epsilon_{\rm s}$, and $\epsilon_{\rm i}$ and the corresponding ranges of the $\epsilon_{\rm c1}$ contributing to each emission component are also indicated in the figure. The dashed green and dotted orange lines denote the b free and $b=1$ fit lines, respectively. As denoted in the figure, the $\epsilon_{\rm c1}$ value is superior to $\epsilon_{\rm cb}$ and $\epsilon_{\rm A}$ values in probing the $\epsilon_{\rm X}$ value.

In Figure~\ref{fig:b_histogram}, we plot the probability density function histogram of the best-fit values of $b$ for the $b$-free models for all the $10^4$ synthetic spectra. The vast majority of the synthetic spectra have a sub-exponential cutoff (i.e., $b<1$) while the distribution peaks at $b\approx 0.85$. We note that this maximum is located at higher $b$ values than those that are usually required to fit the observed phase-averaged spectra of Fermi pulsars (see below the related discussion in Section~\ref{sec:fundamental plane_observations} and Table~\ref{tab:fermi_pulsars}). This  implies that the adopted considerations, which have been broadly deduced from a very small sample of available phase-resolved spectra may not accurately describe the underlying statistics of the Fermi pulsar population. Nonetheless, the resulting $b$ values extend down to $b \approx 0.4$ with $\approx 25\%$ of all the synthetic spectra having $b<0.75$. Thus, below, we examine not only the collective behavior of the entire number of synthetic spectra but also separately the collective behavior of those with $b\lesssim 0.75$.

\begin{figure}[!tb]
\vspace{0.0in}
  \begin{center}
    \includegraphics[width=1.0\linewidth]{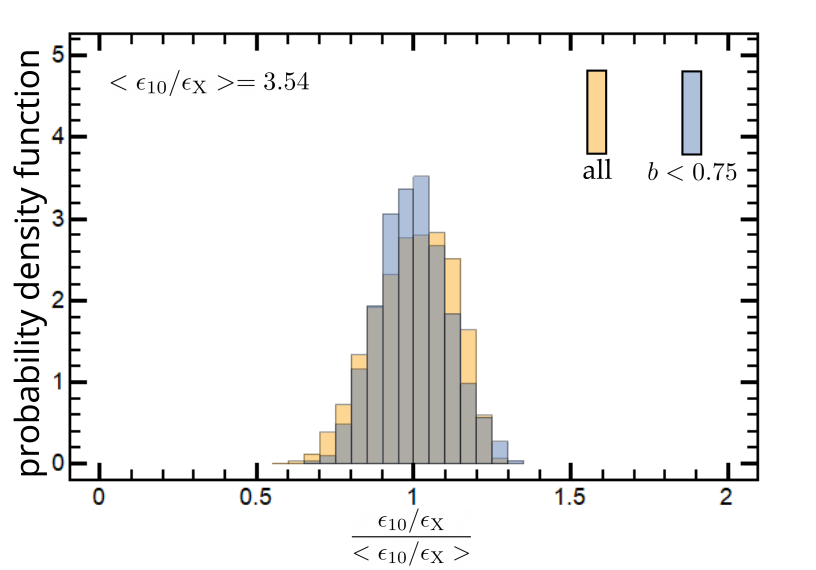}
  \end{center}
  \vspace{-0.2in}
  \caption{Similar to Figure~\ref{fig:ratio_histograms} but for the $\epsilon_{10}$ values.}
  \label{fig:ratio_histograms_e10}
  \vspace{0.0in}
\end{figure}

In Figure~\ref{fig:spectra_examples}, we plot characteristic synthetic phase-averaged spectra with $\epsilon_{\rm X}=3$ GeV and for the indicated $b$ values. The black lines correspond to the synthetic spectra, while the dashed green and the dotted orange lines correspond to the $b$-free and $b$-fixed (i.e., $b=1$) function models, respectively. Similarly to Figure~\ref{fig:synthetic_spectrum_example}, the $\epsilon_{\rm c1}$ is superior in probing the corresponding $\epsilon_{\rm X}$ value.

A characteristic SED energy value would probe the behavior of the cutoff energy value that enters the FP relation (see Eq.~\ref{eq:fp_theory}) if the two energy values are proportional to each other. Thus, the substitution of the cutoff energy value in Eqs.~\eqref{eq:fp_observations}, \eqref{eq:fp_theory} by another characteristic SED energy value that is proportional to it would  reproduce the same exponent possibly affecting only the normalization factor.
In order to see how the various characteristic energy values behave, we calculate for all the synthetic spectra the ratios of these energy values over $\epsilon_{\rm X}$. In Figure~\ref{fig:ratio_histograms}, we plot the probability density histograms of the indicated ratios for the synthetic spectra. We note that in order to make the deviations from the corresponding mean values directly comparable to each other, we have normalized all ratio values to the corresponding mean values, which are also indicated in the panels. The light orange histograms denote the probability densities for all the synthetic spectra while the blue histograms denote only the spectra with $b\leq 0.75$. The $\epsilon_{\rm c1}$ values are not only these that probe better the actual $\epsilon_{\rm X}$ but also those that have the smallest dispersion around the mean ratio value. On the other hand, the $\epsilon_{\rm cb}$ values are those that present the highest spread, having an almost uniform distribution, probing poorly the corresponding $\epsilon_{\rm X}$ values. The $\epsilon_{\rm A}$ values seem to behave fairly even though they are inferior to $\epsilon_{\rm c1}$ in probing $\epsilon_{\rm X}$. More specifically, the mean $\epsilon_{\rm A}/\epsilon_{\rm X}$ ratio deviates considerably from 1 while the ratio values present a wider spread around the corresponding mean value. Moreover, $\epsilon_{\rm A}$ would behave even worse if we had allowed higher $\Gamma$ values because as $\Gamma$ goes close to 2 the corresponding $\epsilon_{\rm A}$ value deviates fast toward very low energy values. On the other hand, the $\epsilon_{\rm c1}$ value is not affected by the $\Gamma$ value. Finally, as we mention in the beginning of this section and as we demonstrate in the next section, the dependence of the spectral photon index on $\ed$ affects $\epsilon_{\rm A}$ introducing biases making $\epsilon_{\rm A}$ practically unsuitable for probing the behavior of the cutoff spectral energy in the FP relation.

\begin{figure}[!tb]
\vspace{0.0in}
  \begin{center}
    \includegraphics[width=1.0\linewidth]{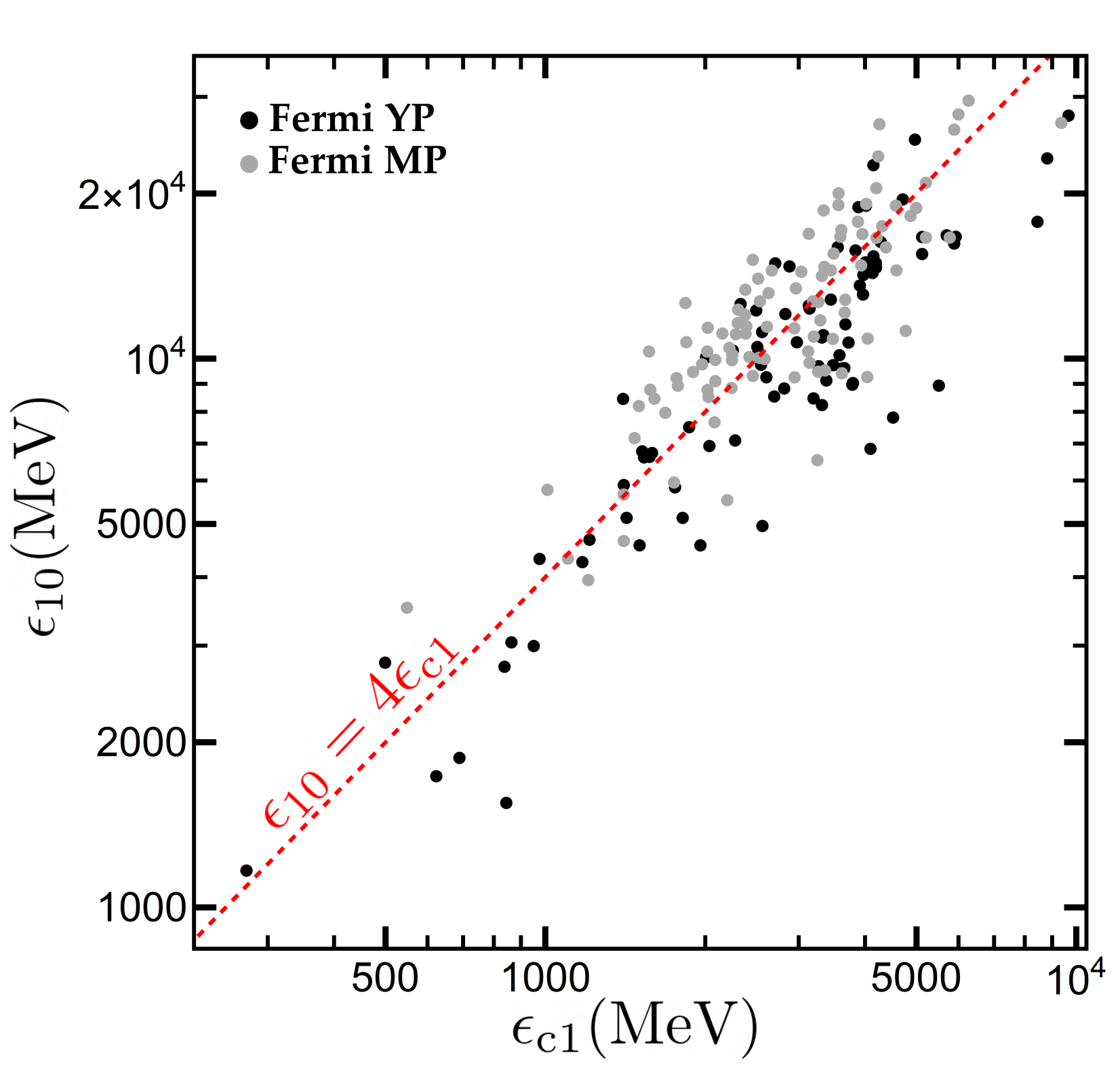}
  \end{center}
  \vspace{-0.2in}
  \caption{The $\epsilon_{10}$ values are plotted as a function of the corresponding $\epsilon_{\rm c1}$ for all the YPs (black points) and MPs (gray points) for which $\epsilon_{\rm A}$ exists (see Table~\ref{tab:fermi_pulsars}). The energies $\epsilon_{10}$ and $\epsilon_{\rm c1}$ are closely proportional. The $\epsilon_{10}=4\epsilon_{\rm c1}$ red dashed line has been plotted to guide the eye.}
  \label{fig:e10vsec1}
  \vspace{0.0in}
\end{figure}

Another way to probe the high-energy part of the phase-averaged spectrum is the identification of the energy level ($\epsilon_{\rm f}>\epsilon_{\rm A}$) where the power of the spectrum reaches a certain fraction, $1/f$, of the SED power corresponding to $\epsilon_{\rm A}$. We have seen that a fraction of one-tenth is adequate because, on the one hand, the corresponding $\epsilon_{\rm 10}$ is not considerably affected by $\epsilon_{\rm A}$ and on the other hand, it is not much higher than $\epsilon_{\rm A}$ and therefore, the flux is still high enough for detection in LAT. In Figure~\ref{fig:en10_demo}, we demonstrate the location of $\epsilon_{\rm 10}$ with respect to $\epsilon_{\rm A}$ in an SED spectrum.
In Figure~\ref{fig:ratio_histograms_e10}, similarly to Figure~\ref{fig:ratio_histograms}, we plot the probability density histograms of the normalized $\epsilon_{10}/\epsilon_{\rm X}$ ratios for the synthetic spectra. We see that the $\epsilon_{10}/\epsilon_{\rm X}$ ratio behaves similarly to the $\epsilon_{\rm c1}/\epsilon_{\rm X}$ ratio and therefore, its use is expected to  be able to reproduce the exponent value of the cutoff energy that enters the FP relation (see Eqs.~\ref{eq:fp_observations} and ~\ref{eq:fp_theory}).

\begin{figure}[!tb]
\vspace{0.0in}
  \begin{center}
    \includegraphics[width=1.0\linewidth]{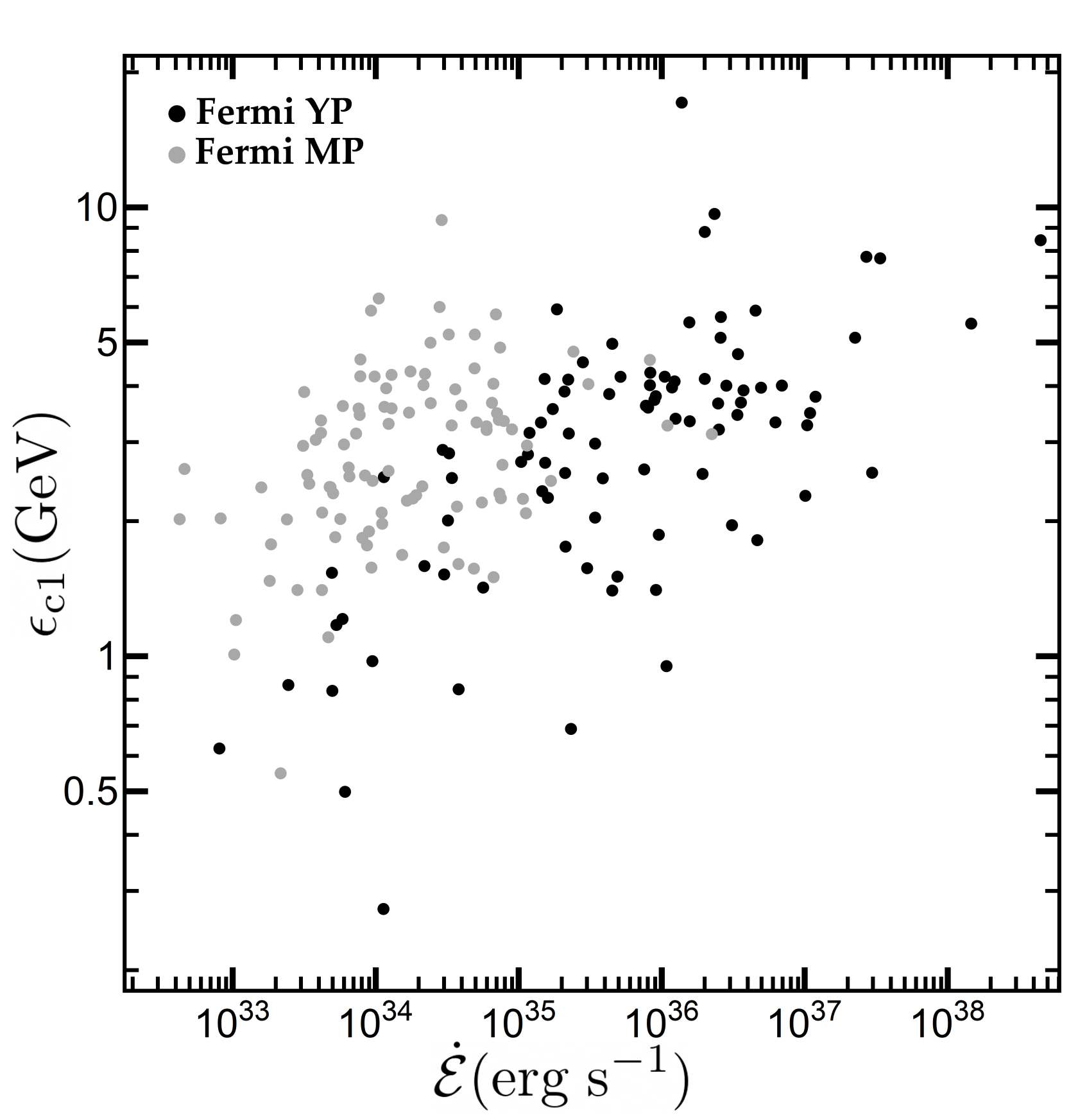}
  \end{center}
  \vspace{-0.2in}
  \caption{The $\epsilon_{\rm c1}$ values are plotted as a function of the corresponding $\ed$ for all the YPs (black points) and MPs (gray points). The energy $\epsilon_{\rm c1}$ shows an increasing trend with $\ed$ for the low $\ed$ values, which, however, seems to saturate toward the high $\ed$ values.}
  \label{fig:ec1vsedot}
  \vspace{0.0in}
\end{figure}

\section{Fundamental Plane: from 4FGL toward 3PC}\label{sec:fundamental plane_observations}

In this section, we update the FP of the observed $\gamma$-ray pulsars taking into account the data analysis of \citetalias{2020ApJS..247...33A}, \citetalias{2022arXiv220111184F}, data from the \citetalias{2005AJ....129.1993M} Pulsar Catalog
as well as the analysis discussed in the previous section.

The calculation of the best-fit parameters for the FP requires the
determination of 4 parameter values (i.e.,
$\ed,~B_{\star},~\ec,~L_{\gamma}$). The \citetalias{2022arXiv220111184F} catalog provides the observed
energy flux, $F_{\rm E}$, above 100 MeV, which is easily converted to $L_{\gamma}$
(i.e., $L_{\gamma}=F_{\rm E}4\pi D^2$) by taking into account the
corresponding object distance $D$, which is taken from the \citetalias{2005AJ....129.1993M} catalog\footnote{We have considered the best estimate distances of the pulsars, i.e., the so-called $Dist$ parameter, provided by the \citetalias{2005AJ....129.1993M} catalog.}. This
conversion assumes, similarly to \citetalias{2013ApJS..208...17A},
that the beaming factor, $f_{\rm b}$, is fixed at 1,
which implies isotropic emission of a certain level.
Furthermore, the \citetalias{2005AJ....129.1993M} catalog provides the $\ed$ value assuming that the
stellar moment of inertia is $I=10^{45}\rm g\;cm^2$. We note that for a number of objects, the \citetalias{2005AJ....129.1993M} catalog provides the Shklovskii $\dot{P}$ corrections. For these objects, we calculate and use the corresponding corrected $\ed$ values. Finally, the \citetalias{2005AJ....129.1993M} catalog does not indicate $\ed$ values for 5 \citetalias{2022arXiv220111184F} objects and therefore, we got these  $\ed$ values from the \citetalias{2013ApJS..208...17A} catalog.

The $\ed$ values also
lead to the polar stellar surface magnetic field values, $B_{\star}=\sqrt{\ed c^3 P^4/4\pi^4 r_{\star}^6 (1+\sin^2
\alpha)}$, assuming the force-free spin-down power \citep{2006ApJ...648L..51S,2009A&A...496..495K,2012MNRAS.424..605P,2013MNRAS.435L...1T} for a magnetic inclination angle, $\alpha=45^{\circ}$, and a stellar radius $r_{\star}=10^6$cm.

The \citetalias{2022arXiv220111184F} catalog also provides the best-fit parameters of the individual phase-averaged spectra using the function model
\begin{equation}
    \label{eq:4FGLspectrumfit}
    \epsilon^2\frac{dN}{d\epsilon}=K\epsilon^2 \left(\frac{\epsilon}{\epsilon_0}\right)^{\frac{d}{b}-\Gamma_{\rm S}}\exp\left[\frac{d}{b^2}\left(1-\left(\frac{\epsilon}{\epsilon_0}\right)^b\right)\right]
    \vspace{0.0in}
\end{equation}
which for $d=0$ takes the form of a pure power law while for $d\neq 0$ describes a family of spectral shapes equivalent to that described by the model of Eq.~\eqref{eq:spectrumfit_bfree}, i.e., power law plus sub-exponential cutoff. Comparing Eqs.~\eqref{eq:spectrumfit_bfree} and \eqref{eq:4FGLspectrumfit}, it becomes evident that
\begin{align}
\label{eq:Gamma_ecb_equivalence}
&\Gamma=\Gamma_{\rm S}-\frac{d}{b}\\
&\epsilon_{\rm cb}=\left(\frac{b^2}{d}\right)^{1/b}\epsilon_0
\end{align}
while the corresponding apex energy of the SED reads
\begin{equation}
\label{eq:eA_equiv}
\epsilon_{\rm A}=\left(\frac{2b+d-b~\Gamma_{\rm S}}{d}\right)^{1/b}\epsilon_0
\end{equation}
which is defined if $d\neq 0$ and $2b+d-b~\Gamma_{\rm S}>0$. Even though Eqs.~\eqref{eq:spectrumfit_bfree} and \eqref{eq:4FGLspectrumfit} can describe, for $d\ne 0$, the same spectral shape families, the form of Eq.~\eqref{eq:4FGLspectrumfit} and the introduction of parameter $d$ provide more reliable fits in the sense that the best-fit parameters do not correlate strongly especially for the objects with low quality data, i.e., less bright objects \citepalias{2022arXiv220111184F}.

Bright pulsars indicate a sub-exponential cutoff (i.e., $b<1$, usually $b\lesssim 0.75$). The function model in Eq.~\eqref{eq:4FGLspectrumfit} allowed fits with free $b$ even for fainter objects (up to 28). However, freeing $b$ for the rest of the objects was not statistically beneficial and therefore, these objects were fitted with $b$ fixed with a value $b=2/3$.

In our study, we use the best-fit parameters provided in the \citetalias{2022arXiv220111184F} catalog to reconstruct the individual
phase-averaged spectra from 0.1 GeV up to 15 GeV. Then, we fit the reconstructed spectra using
the $b=1$ model function that provides the $\epsilon_{\rm c1}$ values that according to our analysis better probe the high-energy part of the spectra and the corresponding $\epsilon_{\rm X}$.

We note that the \citetalias{2022arXiv220111184F} catalog contains 255 pulsars. However, the \citetalias{2005AJ....129.1993M} catalog provides the distance for only 190 of them. Therefore, our analysis incorporates only these 190 pulsars for which the calculation of $L_{\gamma}$ is feasible. For most pulsars (113 out of 190), the derivation of the distance implements the dispersion measure (DM) using the YMW16 model \citep{2017ApJ...835...29Y}. Independent distance estimations are provided for the rest of the pulsars, which are either the only available (for 22 pulsars) or more reliable than the corresponding DM-derived distances (for 55 pulsars). The 22 pulsars without a recorded distance are radio-quiet pulsars since these do not have a DM. The radio-quiet and radio-loud $\gamma$-ray pulsars are indeed expected to trace different combinations of $\alpha$ and $\zeta$, i.e., observer angle. Hence, it becomes evident that the sample of 190 pulsars may trace the ($\alpha,~\zeta$) values corresponding to the radio-loud parameter domain, i.e., small $|\alpha-\zeta |$ values, more efficiently because this domain guarantees the derivation of the distance and, therefore, of $L_{\gamma}$. The $f_{\rm b}$ value depends, in principle, on $\zeta$, which implies that tracing mainly the $\zeta$ values corresponding to those of the radio-loud pulsars could bias the $L_{\gamma}$ values if the $f_{\rm b}$ values of the radio-loud pulsars are systematically different than the average $f_{\rm b}$ values. In any case, the actual $f_{\rm b}$ values remain unknown. In this study, as mentioned above, we assume $f_{\rm b}=1$, which may affect only the normalization factor, since it is a population average and not the exponent values of the FP relation.

In Table~\ref{tab:fermi_pulsars} in the Appendix, we present a set of parameter values, including $\ed,~B_{\star},~\epsilon_{\rm c1},~L_{\gamma}$, for the sample of 190 pulsars included in the \citetalias{2022arXiv220111184F}, which have recorded distances in the \citetalias{2005AJ....129.1993M} catalog.

The luminosity function, i.e., the FP relation,
for the extended sample of the 190 pulsars reads
\begin{equation}
\label{eq:fp_update_ecut1} L_{\gamma}=10^{14.3\pm 1.3}\epsilon_{\rm c1}^{1.39\pm
0.17}B_{\star}^{0.12\pm 0.03}\ed^{0.39\pm 0.05}
\end{equation}
where $\epsilon_{\rm c1}$ is measured in MeV, $B_{\star}$ is measured in G, and $\ed$ and $L_{\gamma}$ are measured in $\rm erg\;s^{-1}$. The expression in Eq.~\eqref{eq:fp_update_ecut1} is consistent with both the FP relation that had taken into account only the \citetalias{2013ApJS..208...17A} data
(see Eq.~\ref{eq:fp_observations}) and the theoretically predicted one
(see Eq.~\ref{eq:fp_theory}). We also derived the FP relation $L_{\gamma}=10^{18.1\pm 2.6}\epsilon_{\rm c1}^{1.17\pm 0.76}B_{\star}^{0.13\pm 0.09}\ed^{0.30\pm 0.14}$ corresponding only to the 22 radio-quiet pulsars, which, despite the large uncertainties that are mainly due to the small number statistics, is compatible with the Eq.~\ref{eq:fp_update_ecut1} implying that the FP relation in Eq.~\ref{eq:fp_update_ecut1} is not drastically affected by this selection bias.

As mentioned above, the $\epsilon_{10}$ values also probe the high-energy parts of the
spectra. Thus, the FP relation or luminosity function that incorporates the corresponding $\epsilon_{10}$ values reads
\begin{equation}
\label{eq:fp_update_ef1/10} L_{\gamma}=10^{10.2\pm 1.35} \epsilon_{10}^{1.22\pm
0.17}B_{\star}^{0.11\pm 0.03}\ed^{0.51\pm 0.05}
\end{equation}
which shows that the exponents are consistent with those in all the
previous FP versions supporting the FP underlying theoretical considerations.
We also note that the uncertainties in the above equations, as well as those in Eq.~\ref{eq:fp_observations}, are purely statistical, and do not explicitly incorporate systematic uncertainties associated with distances or other variables; however, these systematics may implicitly affect the quoted statistical uncertainty in the luminosity function exponents).

The exponent consistency between the FP
relations in Eqs.~\eqref{eq:fp_update_ecut1} and \eqref{eq:fp_update_ef1/10} implies that
the ratio $\epsilon_{\rm c1}/\epsilon_{\rm 10}$ is practically constant (i.e., proportional relationship between $\epsilon_{\rm c1}$ and $\epsilon_{10}$). Indeed, as shown in Figure~\ref{fig:e10vsec1}, where we plot $\epsilon_{10}$ as a function of $\epsilon_{\rm c1}$, $\epsilon_{\rm 10}$ is closely proportional to $\epsilon_{\rm c1}$. We also note that $\epsilon_{10}$ and $\epsilon_{\rm c1}$ are not directly related through the same function in the sense that the $\epsilon_{10}$ value is derived from best fit corresponding to the model function in Eq.~\eqref{eq:4FGLspectrumfit} while the $\epsilon_{\rm c1}$ value is the best-fit parameter of the function model in Eq.~\eqref{eq:spectrumfit_bfree} for $b=1$.

In Figure~\ref{fig:ec1vsedot}, we plot the $\epsilon_{\rm c1}$ values as a function of the corresponding $\ed$ values, which shows a weakly increasing trend of $\epsilon_{\rm c1}$ with $\ed$.

\begin{figure}[!tb]
\vspace{0.0in}
  \begin{center}
    \includegraphics[width=1.0\linewidth]{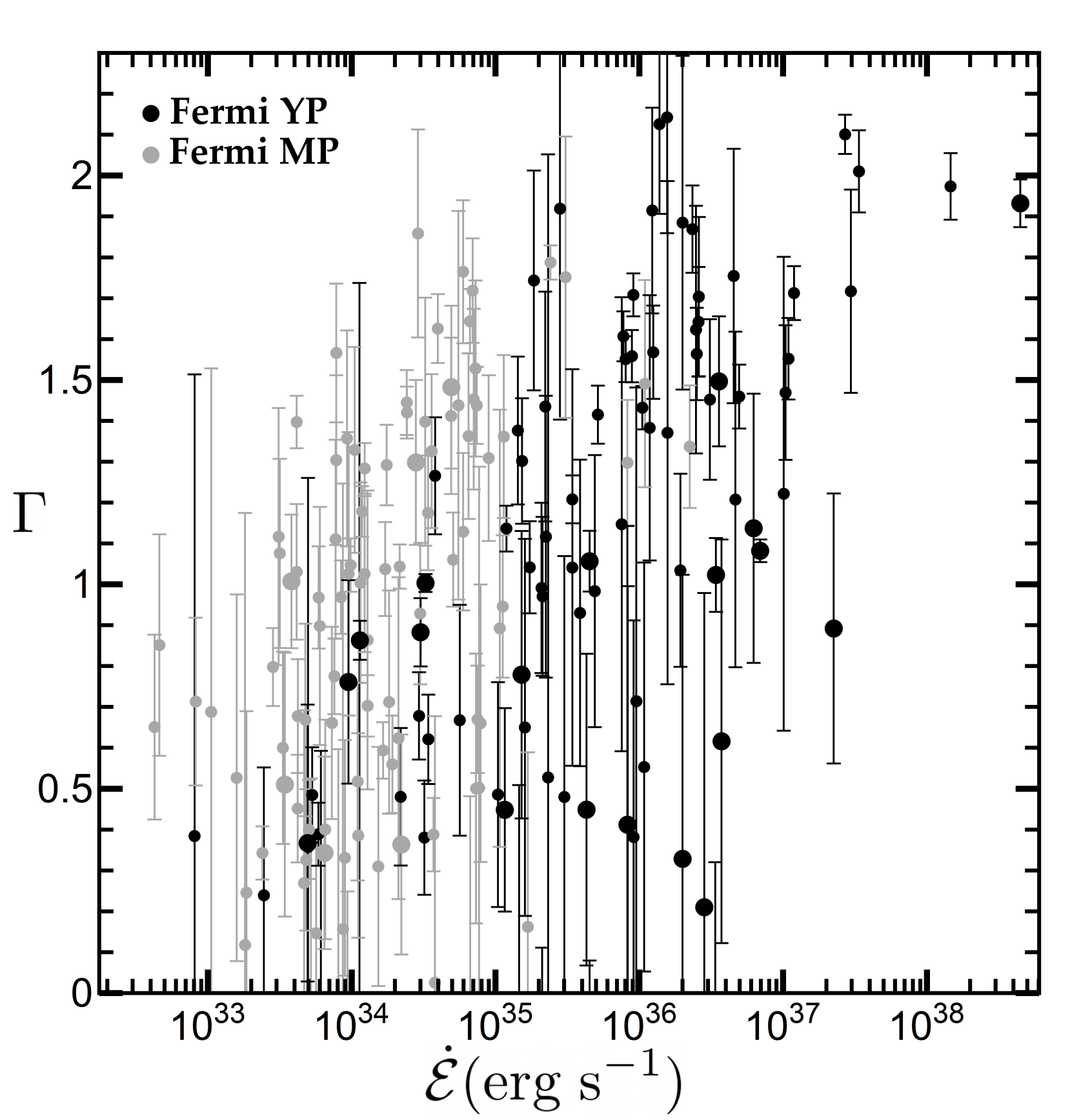}
  \end{center}
  \vspace{-0.2in}
  \caption{The spectral photon index $\Gamma$ is plotted as a function of $\ed$ for all the YPs (black points) and MPs (gray points). The $\Gamma$ values have been calculated using Eq.~\ref{eq:Gamma_ecb_equivalence}, while the corresponding errors have been calculated considering the error propagation and the values and errors of $\Gamma_{\rm S},~d,~b$, provided in \citetalias{2022arXiv220111184F}. The big dots denote the high-significance sources, i.e., $b$-free sources. The spectral index shows an increasing trend with $\ed$ despite the large spread. Moreover, the MPs seem to have higher $\Gamma$ values compared to those of YPs, for the same $\ed$ values.}
  \label{fig:Gammavsedot}
  \vspace{0.0in}
\end{figure}

\begin{figure}[!tb]
\vspace{0.0in}
  \begin{center}
    \includegraphics[width=1.0\linewidth]{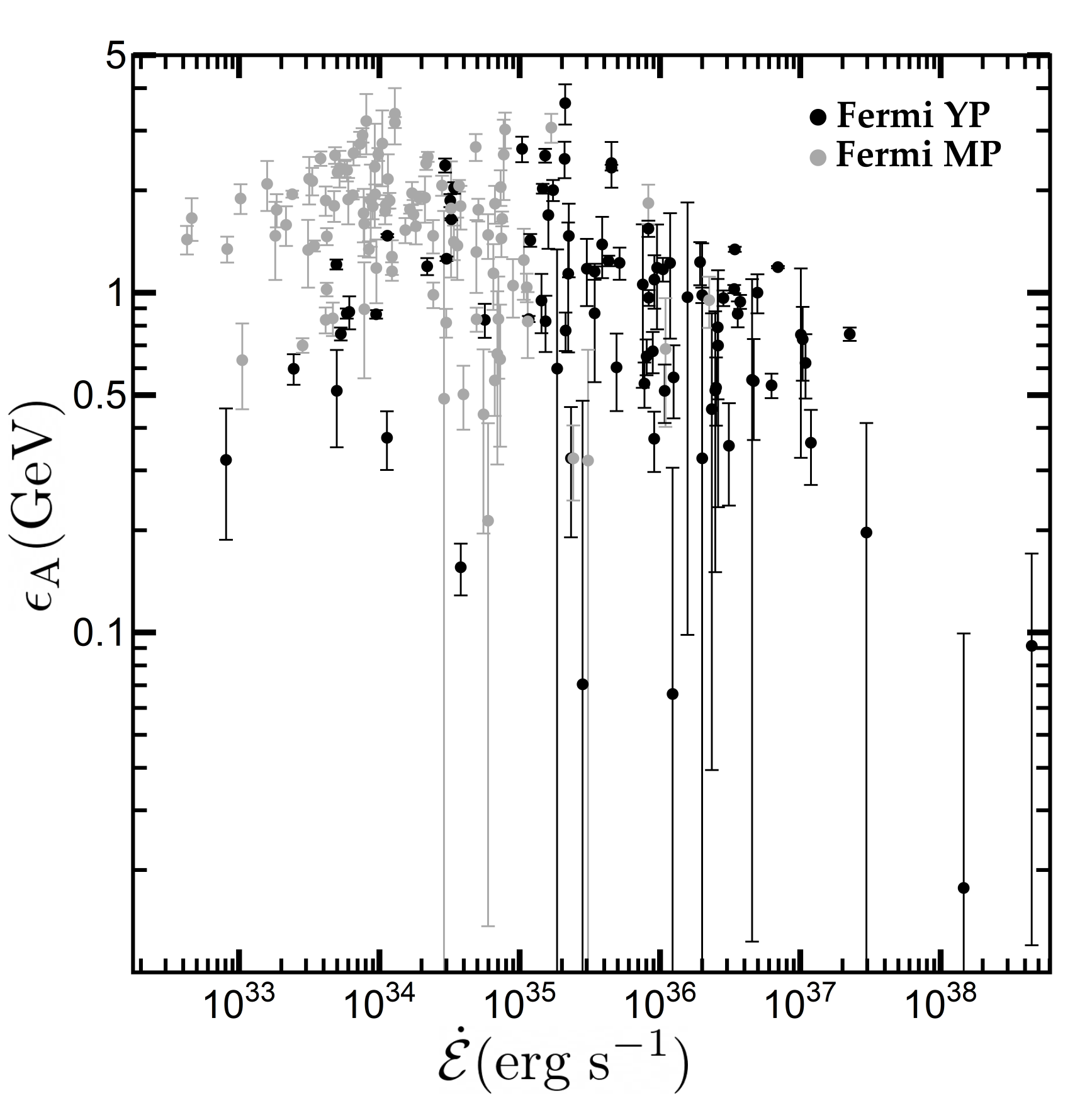}
  \end{center}
  \vspace{-0.2in}
  \caption{The apex energy, $\epsilon_{\rm A}$, of the SED spectra is plotted as a function of $\ed$ for all the YPs (black points) and MPs (gray points). The behavior of $\epsilon_{\rm A}$ is affected by the behavior of $\Gamma$. The plotted errors have been taken from \citetalias{2022arXiv220111184F}. The increasing $\Gamma$ toward high $\ed$ affects the behavior of $\epsilon_{\rm A}$ (compared to that of $\epsilon_{\rm c1}$), which especially for $\ed\gtrsim 10^{34}\rm erg\;s^{-1}$ show a decreasing trend with $\ed$. We note that the points with the largest errors that appear mostly for the low $\epsilon_{\rm A}$ values correspond to pulsars with $\Gamma$ values close to 2.}
  \label{fig:eAvsedot}
  \vspace{0.0in}
\end{figure}

\begin{figure*}[t]
\vspace{0.0in}
  \begin{center}
    \includegraphics[width=1.0\linewidth]{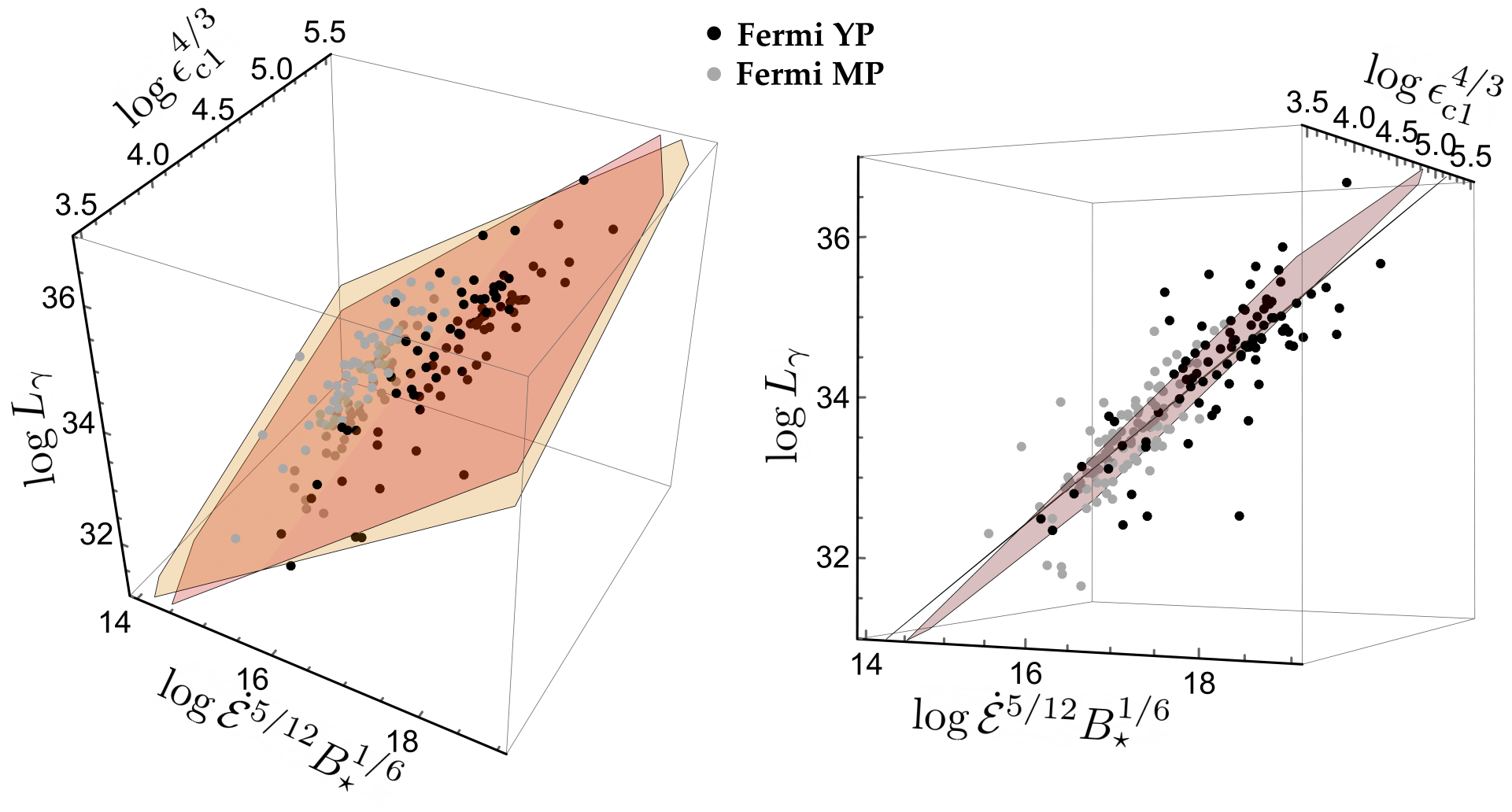}
  \end{center}
  \vspace{-0.2in}
  \caption{The FP plotted in the reduced dimensinality space $(\mathcal{X},~\mathcal{Y},~\mathcal{Z})=(\ed^{5/12}B_{\star}^{1/6},~\epsilon_{\rm c1}^{4/3},~L_{\gamma})$. The data points correspond to the YPs (black color) and MPs (gray color), while the reddish plane denotes the theoretical FP (i.e., $\mathcal{Z}\propto\mathcal{X}\mathcal{Y}$; see Eq.~\ref{eq:fp_theory}) proposed by \citetalias{2019ApJ...883L...4K} and the yellowish plane denotes the the best fit of the observed Fermi pulsars. In the right-hand panel the yellowish plane is shown edge on.}
  \label{fig:FP4FGL}
  \vspace{0.0in}
\end{figure*}

In \S~\ref{sec:spectrum_function_model}, we discussed that, in general, $\epsilon_{\rm A}$ is expected to probe relatively well $\epsilon_{\rm X}$ even though its performance is inferior to those of $\epsilon_{\rm c1}$ and $\epsilon_{10}$. As also mentioned in \S~\ref{sec:intro} and \S~\ref{sec:spectrum_function_model}, the greatest issue with $\epsilon_{\rm A}$ is that it depends, in principle, on the the spectral photon index, $\Gamma$ (see Eqs.~\ref{eq:spectrumfit_bfree} and \ref{eq:Gamma_ecb_equivalence}). However, as demonstrated in Figure~\ref{fig:Gammavsedot}, where we plot $\Gamma$ as a function of $\ed$, the observed $\Gamma$ values show an increase with $\ed$ \citepalias[see also][]{2013ApJS..208...17A}, which affects the behavior of $\epsilon_{\rm A}$. The increase of $\Gamma$ pushes $\epsilon_{\rm A}$ down (see Eq.~\ref{eq:apex_energy}). We note that the $\Gamma$ values have a large scatter despite the increasing trend with $\ed$. Moreover, the apparent scatter is further enhanced by the presence of the high significance sources, i.e., those with $b$ free (denoted by the big dots in Figure~\ref{fig:Gammavsedot}), which are mainly located below the main diagonal.

In Figure~\ref{fig:eAvsedot}, which shows $\epsilon_{\rm A}$ as a function of $\ed$, we see a behavior that actually reflects the behavior of $\Gamma$ with $\ed$. More specifically, the $\epsilon_{\rm A}$ values for the high $\ed$ values, where $\Gamma$ is higher, are considerably lower than the corresponding $\epsilon_{\rm c1}$ values (compared to the situation at low $\ed$ values). As a consequence  $\epsilon_{\rm A}$ seem stabilized or even decreasing with $\ed$.

The FP denotes a 3-dimensional (3D) plane embedded in the 4-dimensional (4D)
logarithmic space of ($\ed,~B_{\star},~\epsilon_{\rm c1},~L_{\gamma}$) and therefore, it
is difficult to be visualized. In order to provide a visualization of the FP,
we present the FP and the corresponding data in a compacted 3D space. This
space involves the variables
$(\mathcal{X},~\mathcal{Y},~\mathcal{Z})=(\ed^{5/12}B_{\star}^{1/6},~\epsilon_{\rm c1}^{4/3},~L_{\gamma})$.
In these variables the theoretical FP relation (Eq.~\ref{eq:fp_theory}) reads
$\mathcal{Z}\propto\mathcal{X}\;\mathcal{Y}$.

In Figure~\ref{fig:FP4FGL}, we plot the 190 pulsars points in the compacted
$(\mathcal{X},~\mathcal{Y},~\mathcal{Z})$ space seen from two different points of view. The transparent reddish plane corresponds to the
theoretical FP (i.e., $\mathcal{Z}\propto\mathcal{X}\;\mathcal{Y}$) while the
transparent yellowish plane corresponds to the fitting of the observed points. An edge-on view of the fitting yellowish plane is shown in the right-hand side panel. We note that the scattering around the FP has a standard deviation $\sim 0.33$ dex, similar to the one found in \citetalias{2019ApJ...883L...4K}. Nonetheless, the errors of the fitting parameters are reduced compared to those corresponding to the fitting of the significant smaller sample of objects compiled in \citetalias{2019ApJ...883L...4K}.

\section{The Fundamental Plane Revisited}\label{sec:FPrevisited}

In \citetalias{2019ApJ...883L...4K}, one of the main assumptions that led to the theoretical FP relation (i.e., Eq.~\ref{eq:fp_theory}) was that the CR is in the RRL regime. Nonetheless, a more careful investigation shows that the RRL regime requirement is sufficient but not necessary. The RRL regime balances the rate of energy gain to the rate of radiation energy losses. Setting the region of interest near the LC, the RRL regime simply means that
\begin{equation}
\label{eq:RRLR} E_{\rm BLC}B_{\rm LC}\propto \gamma_{\rm L}^4R_{\rm C}^{-2}
\end{equation}
where $E_{\rm BLC}$ is the accelerating electric field in $B_{\rm LC}$ units while $\gamma_{\rm L}$ and $R_{\rm C}$ denote the Lorentz factor, i.e., particle energy, and radius of curvature, respectively. In \citetalias{2019ApJ...883L...4K}, we claimed that the $\gamma$-ray luminosity of an emitting particle scales as $E_{\rm BLC}B_{\rm LC}$ and therefore as $\gamma_{\rm L}^4R_{\rm C}^{-2}$, which is true in the RRL regime. However, the emitting power scales always as $\gamma_{\rm L}^4 R_{\rm C}^{-2}$ independently of whether this power matches the accelerating one or not. It is the $\gamma_{\rm L}^4 R_{\rm C}^{-2}$ scaling that leads to the FP relation \citepalias{2019ApJ...883L...4K}, which makes the RRL regime requirement unnecessary. Nonetheless, the RRL regime assumption allows the determination of the accelerating electric field component (i.e., $E_{\rm BLC}$), connecting it monotonically to the observed $\gamma$-ray efficiency.

\begin{figure}[!tb]
\vspace{0.0in}
  \begin{center}
    \includegraphics[width=1.0\linewidth]{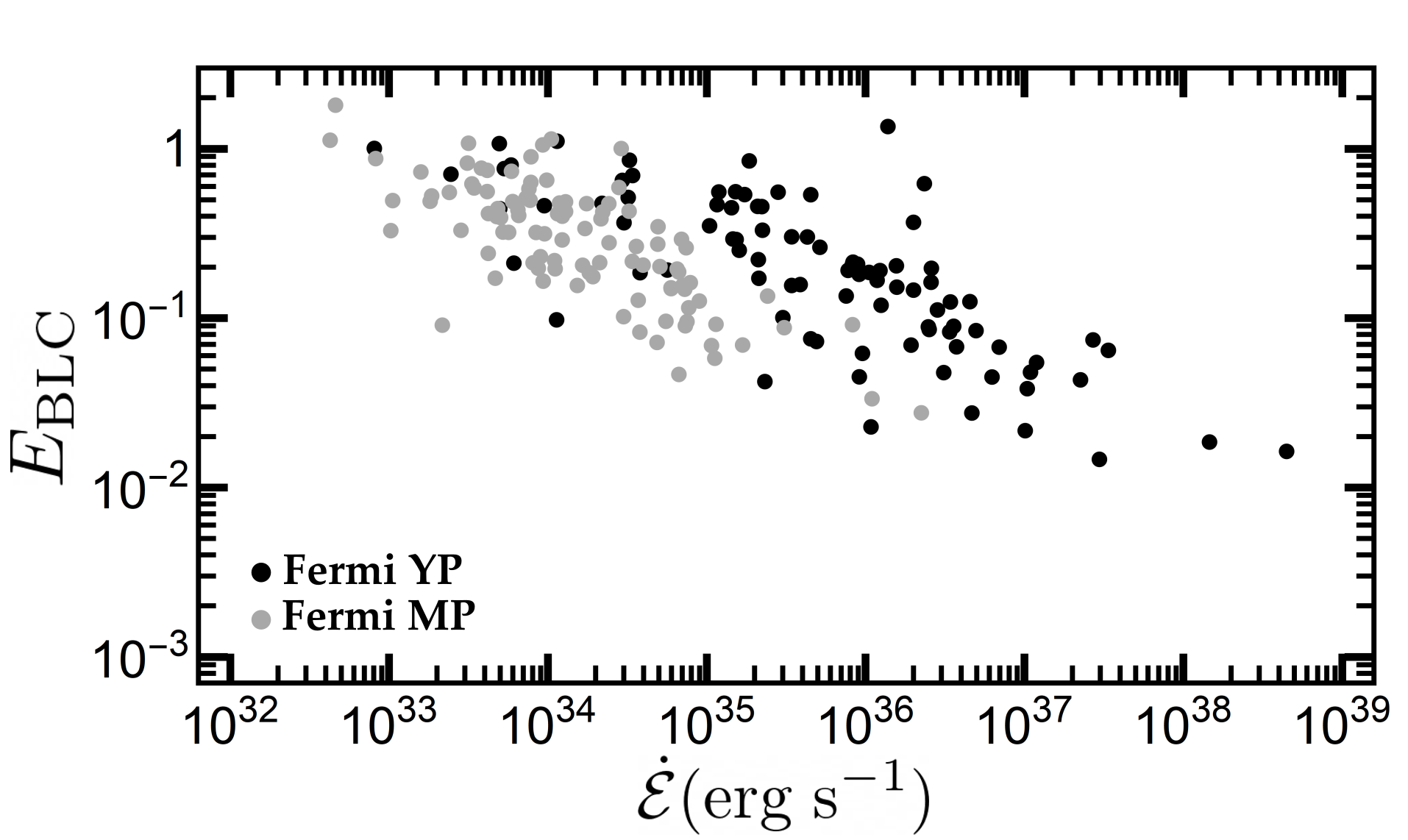}
  \end{center}
  \vspace{-0.2in}
  \caption{The FP does not require CR emission in the RRL regime. However, assuming CR emission in the RRL regime, the accelerating electric field component, $E_{\rm BLC}$, in $B_{\rm LC}$ units, is revealed. Here, we plot $E_{\rm BLC}$ as a function of $\ed$ for all the YPs (black points) and MPs (gray points). The derivation of $E_{\rm BLC}$ assumes that $R_{\rm C}\approx R_{\rm LC}$ and $r_{\star}=10^6$~cm (see Eq.~\ref{eq:EBLC}).}
  \label{fig:eaccvsedot}
  \vspace{0.0in}
\end{figure}

In \citet[][see also \citetalias{2019ApJ...883L...4K}]{2017ApJ...842...80K}, assuming the RRL regime and emission near the LC, we derived the accelerating electric field component and its behavior with $\ed$ that revealed a decreasing behavior of $E_{\rm BLC}$ with $\ed$ at high $\ed$ values and a saturation trend toward low $\ed$ values indicating a fixed dissipation power toward low $\ed$ and a decreasing dissipation power toward high $\ed$. Figure~\ref{fig:eaccvsedot}, which presents $E_{\rm BLC}$ as a function of $\ed$ is an update of the bottom panel of figure 2 in \citet{2017ApJ...842...80K} that had used the sample and data analysis of \citetalias{2013ApJS..208...17A}. We briefly remind the reader that the expression of the cutoff energy, i.e., $\epsilon_{\rm c1}=3 c \hbar \gamma_{\rm L}^3/2R_{\rm C}$ provides an estimation of the characteristic $\gamma_{\rm L}$ value for each observed $\epsilon_{\rm c1}$ value assuming that $R_{\rm C}\approx R_{\rm LC}$. Then taking into account these $\gamma_{\rm L}$ and $R_{\rm C}$ values and assuming the RRL regime $E_{\rm BLC}$ reads \citep[see also][]{2017ApJ...842...80K}.
\begin{equation}
\label{eq:EBLC} E_{\rm BLC}=\frac{E_{\rm acc}}{B_{\rm LC}}=\frac{c}{(3\pi)^{7/3}r_{\star}^3 \hbar^{4/3}}
\frac{P^{7/3}\epsilon_{\rm c1}^{4/3}}{B_{\star}}
\end{equation}
where $E_{\rm acc}$ is the accelerating electric field component. We also note that in Eq.~\eqref{eq:EBLC}, the radius of curvature has been considered to be equal to $R_{\rm LC}$\footnote{It is noted that \citet{2012ApJ...754...33L} calculated for 46 pulsars an $\epsilon_{\rm cut}$ assuming $E_{\rm BLC}=1$ and $R_{\rm C}=R_{\rm LC}$ and compare them to the corresponding observed values. This comparison implied that $E_{\rm BLC}\lesssim 1$ for the vast majority of pulsars. However, neither the $E_{\rm BLC}$ values nor their behavior with $\ed$ was reported.}.

\section{Discussion and Conclusions}\label{sec:conclusions}

The discovery of the FP relation and the underlying simple theoretical description of it \citepalias{2019ApJ...883L...4K} set strong diagnostics for the modeling of the magnetosphere structure and pulsar high-energy emission. Thus, a major goal of our study was the exploration of the robustness of the FP relation using a considerably larger data-set that has become available through the recent publication of the \citetalias{2022arXiv220111184F} catalog. The motivation behind this exploration was the development of advanced kinetic particle-in-cell models that will be compared to the data evaluating with confidence their ability to reproduce the observed trends (e.g., the FP relation). We have already developed these kind of models and the results of this study will be presented in a forthcoming paper.

The FP relation incorporates four parameters, i.e., $\ed, B_{\star}, \epsilon_{\rm cut}, L_{\gamma}$. The former two are easily and more or less unambiguously derived by the the observed $P, \dot{P}$ values even though their derivation requires some additional assumptions regarding the moment of inertia value(s) and the spin-down power law (e.g., vacuum, force-free).

An accurate derivation of $L_{\gamma}$ is more problematic not only because of the related flux and distance uncertainties\footnote{The flux and distance uncertainties contribute to the spread around the FP leaving, however, unaffected the scalings, i.e., the exponents in the FP relation.}, but also because it depends on the beaming factor, which measures the uniformity of emission on the sky \citepalias{2013ApJS..208...17A}. The beaming factor varies considerably between different models while it may also vary within the same model depending, in principle, on the observer angle. However, the consideration of a different average beaming-factor value would collectively affect all the $L_{\gamma}$ values leaving unaffected the corresponding scaling, i.e., the exponents in the FP relation. Any $L_{\gamma}$ ambiguity that is the result of the variation of beaming factors around the mean value contributes just to the spread around the FP.

On the other hand, the derivation of the characteristic $\epsilon_{\rm cut}$ value is the most intriguing one for a variety of reasons. The exponent of $\epsilon_{\rm cut}$ in the FP relation is larger than all the other exponents, which implies that the luminosity function is more sensitive to $\epsilon_{\rm cut}$. Moreover, the observations indicate that the spectral characteristic $\epsilon_{\rm cut}$ values have a rather limited range that barely exceeds one order of magnitude. Finally, the observed phase-averaged spectral shapes and the function models that seem to fit them provide a variety of characteristic energies whose exact physical meaning and their possible relation to the synchro-curvature emission, near the curvature radiation regime, is not always clear.

The phase-averaged spectra consist of collections of emission from different magnetosphere regions reflecting, in general, different physical properties (e.g., accelerating electric field components, particle energies, radii of curvature of particle trajectories). The phase-resolved spectra corresponding to different phases may also trace different magnetosphere regions although, in this case, the variety of the physical properties should be significantly reduced compared to that in the phase-averaged spectra.

The most relevant $\epsilon_{\rm cut}$ values that should enter the FP relation are the ones corresponding to the peaks of the $\gamma$-ray light curves, which trace the $\gamma$-ray emission corresponding to the core of the dissipative regions, i.e., the equatorial current sheet. However, despite the existence of tens of bright pulsars, only a handful of them have published phase-resolved spectra, which makes the phase-averaged spectra currently the only available source of analysis.

Single $e^{\pm}$ radiating in single region of space produce spectra with exponential cutoffs, while the phase-averaged spectra are better fitted by sub-exponential cutoffs. The contribution of different magnetosphere regions is able to explain the significant deviation from the pure exponential cutoff. Thus, we decided to use a toy-model approach to explore the phase-averaged spectra behavior assuming that these are simply superpositions of pure exponential cutoff spectra (i.e., the phase-resolved ones). Our considerations for constructing synthetic spectra were based on the properties of the few available phase-resolved spectra. However, these considerations taken from only a few cases were quite broad and therefore, they were probably not able to correctly describe the spectral shape statistics of the entire Fermi population. Nonetheless, our goal was not to reproduce the spectral statistics of the observed objects, but to explore how well various energy parameters probe the highest  characteristic $\epsilon_{\rm cut}$ energy value between the two peaks, i.e., the $\epsilon_{\rm X}$ value that should be considered in the FP relation.

Thus, we examined different spectrum function models, focusing on the determination of the capacity of the various spectral photon energy parameters for probing $\epsilon_{\rm X}$. Using the synthetic phase-averaged spectra, we found that indeed these have sub-exponential cutoffs. However, we showed that the cutoff energy corresponding to the pure exponential cutoff function model is superior in probing $\epsilon_{\rm X}$. Thus, our analysis implies that even though the sub-exponential model function may better describe the totality of the spectra, the cutoff energy that is more relevant to the $\ec$ in the FP relation is better captured by the one corresponding to the pure exponential model function, i.e., $\epsilon_{\rm c1}$.

We note that the shape of most of the synthetic spectra showed a sub-exponential cutoff corresponding to a $b$ exponent (see Eq.~\ref{eq:spectrumfit_bfree}) that was higher than those indicated by the most bright pulsars even though the fitting of several synthetic spectra provided quite low $b$ values. The existence of the relatively high $b$ values implies that either the spectral rules (e.g., relative powers and characteristic cutoff energies of the main emission components) adopted in our study do not always describe realistic spectra or the individual spectra that contribute to the synthetic phase-averaged ones are not purely exponential. As discussed above, the phase-resolved spectra may also probe different magnetosphere regions that contribute to the same phase \citep{2010ApJ...715.1282B,2010MNRAS.404..767C,2014ApJ...793...97K}. Nonetheless, the variety of the physical properties (e.g., $\gamma_{\rm L}$, accelerating field components, $R_{\rm C}$) of the regions contributing only to one phase is expected to be smaller than those in the phase-averaged spectra. This practically means that any deviation of the phase-resolved spectra from the pure exponential cutoff should be smaller (i.e., $b$ closer to 1) than what is observed in the phase-averaged spectra. A contribution of phase-resolved spectra with smaller (than 1) $b$ values would result in phase-averaged spectra with even smaller $b$ values. Finally, there is a possibility that the $b$ value at least for some pulsars is affected by the existence of an additional very high energy emission component \citep[inverse Compton, IC; see][]{2021ApJ...923..194H}, which may form a softer transition between the two spectrum components (i.e., CR and IC).

In any case, a detailed description of the variety of the phase-resolved spectra will remain unclear until we get access to a larger amount of high-quality data. Moreover, any, so-far, available phase-resolved spectra have adopted the pure exponential model function, i.e., $b=1$ \citep[see][]{2013PhDT.......182D}. Hopefully, this work will further motivate observers to perform a detailed phase-resolved spectroscopy incorporating, if possible, more objects.

Taking into account our conclusions regarding the interpretation of $\epsilon_{\rm c1}$, we used the results of the spectral analysis in \citetalias{2022arXiv220111184F} to reconstruct the phase-averaged spectra for 190 pulsars, which allowed the calculation of the corresponding $\epsilon_{\rm c1}$ values. Taking also into account the \citetalias{2022arXiv220111184F} $\gamma$-ray fluxes as well as the distances and the spin-down powers from the \citetalias{2005AJ....129.1993M} catalog, we updated the observed FP relation by increasing more than two fold the number of Fermi pulsars compared to those we used in our original study \citetalias{2019ApJ...883L...4K}. The extended sample showed results in agreement with those corresponding to the sample of \citetalias{2013ApJS..208...17A} indicating that the high-energy part of the Fermi spectra is consistent with CR.

Our study provides a preamble for the anticipated 3PC focusing mainly on the exploitation and further interpretation of the \citetalias{2022arXiv220111184F} data that allow the derivation of meaningful parameter values that enter the FP relation. Nonetheless, the 3PC is expected to provide advanced data treatments and possibly improved fitting function models expanding also considerably the number of objects compiled in this study.

Finally, we revisited the FP theoretical background by relaxing the requirements that lead to the derivation of the theoretical FP relation. More specifically, we showed that the RRL regime assumption is redundant for the FP relation; however, it sets additional constraints that allow the determination of the underlying accelerating electric field component.

\acknowledgements We acknowledge helpful discussions with Isabelle Grenier and David Smith. We would like to thank the LAT group for providing valuable comments and suggestions, which helped us to improve the quality of the paper. We would also like to thank the International Space Science Institute (ISSI) for providing financial support for the organization of the meeting of the ISSI Team that was led by I. Contopoulos and D. Kazanas. C.K. and Z.W. are supported by the
Fermi Guest Investigator program and the NASA Theory Program. D. K. is supported by the NASA Data Analysis Program. The material is based upon work supported by
NASA under award number 80GSFC21M0002.

\appendix
\label{sec:appendix}

\startlongtable
\begin{deluxetable*}{lclrrcrrrrrrr}
\label{tab:fermi_pulsars}




\tablecaption{A set of parameters for 190 \citetalias{2022arXiv220111184F} young and millisecond pulsars.}



\tablehead{\colhead{PSR} & \colhead{$\ed$} & \colhead{$B_{\star}$} & \colhead{P} & \colhead{$\epsilon_{c1}$} & \colhead{$L_{\gamma}$} & \colhead{$\epsilon_{\rm A}$} & \colhead{$\epsilon_{cb}$} & \colhead{$\epsilon_{10}$} & \colhead{$\Gamma_{\rm S}$} & \colhead{$\epsilon_0$} & \colhead{$d$} & \colhead{$b$} \\
\colhead{} & \colhead{$(\rm erg\;s^{-1})$} & \colhead{(G)} & \colhead{(ms)} & \colhead{(GeV)} & \colhead{$(\rm erg\;s^{-1})$} & \colhead{(GeV)} & \colhead{(GeV)} & \colhead{(GeV)} & \colhead{} & \colhead{(GeV)} & \colhead{} & \colhead{} }

\startdata
$\rm J0002+6216$ & $1.5\times 10^{35}$ & $1.1\times 10^{12}$ & $115.4$ & $2.70$ & $9.1\times 10^{34}$ & $0.82$ & $0.77$ & $8.53$ & $2.25$ & $1.31$ & $0.633$ & $0.667$ \\
$\rm J0007+7303$ & $4.5\times 10^{35}$ & $1.4\times 10^{13}$ & $315.9$ & $4.97$ & $1.0\times 10^{35}$ & $2.33$ & $0.83$ & $25.10$ & $1.97$ & $2.21$ & $0.495$ & $0.540$ \\
$\rm J0023+0923$ & $1.2\times 10^{34}$ & $2.2\times 10^8$ & $3.1$ & $2.59$ & $3.1\times 10^{33}$ & $1.28$ & $0.72$ & $9.99$ & $1.83$ & $0.95$ & $0.535$ & $0.667$ \\
$\rm J0030+0451$ & $3.4\times 10^{33}$ & $3.0\times 10^8$ & $4.9$ & $2.42$ & $7.6\times 10^{32}$ & $1.37$ & $0.19$ & $10.10$ & $1.99$ & $1.36$ & $0.784$ & $0.529$ \\
$\rm J0034-0534$ & $2.4\times 10^{34}$ & $1.2\times 10^8$ & $1.9$ & $3.66$ & $4.4\times 10^{33}$ & $0.99$ & $1.22$ & $12.10$ & $1.91$ & $0.76$ & $0.326$ & $0.667$ \\
$\rm J0101-6422$ & $8.4\times 10^{33}$ & $1.3\times 10^8$ & $2.6$ & $2.53$ & $1.6\times 10^{33}$ & $1.34$ & $0.70$ & $10.00$ & $1.76$ & $0.90$ & $0.528$ & $0.667$ \\
$\rm J0102+4839^*$ & $1.8\times 10^{34}$ & $2.5\times 10^8$ & $3.0$ & $4.31$ & $9.6\times 10^{33}$ & $1.70$ & $1.55$ & $17.40$ & $1.91$ & $1.38$ & $0.410$ & $0.667$ \\
$\rm J0106+4855$ & $2.9\times 10^{34}$ & $2.5\times 10^{11}$ & $83.2$ & $2.88$ & $2.2\times 10^{34}$ & $2.37$ & $0.85$ & $14.70$ & $1.69$ & $1.58$ & $0.672$ & $0.667$ \\
$\rm J0205+6449$ & $2.7\times 10^{37}$ & $4.8\times 10^{12}$ & $65.7$ & $7.76$ & $7.9\times 10^{34}$ & - & $3.75$ & - & $2.35$ & $0.86$ & $0.167$ & $0.667$ \\
$\rm J0218+4232$ & $2.4\times 10^{35}$ & $5.7\times 10^8$ & $2.3$ & $4.77$ & $5.7\times 10^{34}$ & $0.33$ & $1.81$ & $11.20$ & $2.18$ & $0.82$ & $0.262$ & $0.667$ \\
$\rm J0248+4230$ & $3.8\times 10^{34}$ & $2.8\times 10^8$ & $2.6$ & $1.60$ & $7.9\times 10^{32}$ & $1.80$ & $0.35$ & $8.46$ & $1.74$ & $1.45$ & $1.140$ & $0.667$ \\
$\rm J0248+6021$ & $2.1\times 10^{35}$ & $4.7\times 10^{12}$ & $217.1$ & $1.75$ & $1.4\times 10^{34}$ & $0.77$ & $0.40$ & $5.83$ & $2.28$ & $1.11$ & $0.871$ & $0.667$ \\
$\rm J0251+2606$ & $1.8\times 10^{34}$ & $1.9\times 10^8$ & $2.5$ & $2.25$ & $8.0\times 10^{32}$ & $1.57$ & $0.58$ & $9.94$ & $1.82$ & $1.24$ & $0.736$ & $0.667$ \\
$\rm J0312-0921$ & $1.5\times 10^{34}$ & $3.6\times 10^8$ & $3.7$ & $1.68$ & $4.4\times 10^{32}$ & $1.53$ & $0.38$ & $7.97$ & $1.71$ & $1.15$ & $0.933$ & $0.667$ \\
$\rm J0340+4130$ & $7.6\times 10^{33}$ & $2.0\times 10^8$ & $3.3$ & $3.56$ & $5.9\times 10^{33}$ & $2.91$ & $1.17$ & $19.10$ & $1.62$ & $1.66$ & $0.562$ & $0.667$ \\
$\rm J0357+3205$ & $5.9\times 10^{33}$ & $3.2\times 10^{12}$ & $444.1$ & $1.21$ & $5.0\times 10^{33}$ & $0.87$ & $0.23$ & $4.68$ & $2.28$ & $1.10$ & $1.260$ & $0.667$ \\
$\rm J0437-4715$ & $2.8\times 10^{33}$ & $3.8\times 10^8$ & $5.8$ & $1.40$ & $5.2\times 10^{31}$ & $0.70$ & $0.29$ & $4.65$ & $1.90$ & $0.61$ & $0.732$ & $0.667$ \\
$\rm J0514-4408$ & $2.5\times 10^{33}$ & $1.1\times 10^{12}$ & $320.3$ & $0.86$ & $6.2\times 10^{32}$ & $0.60$ & $0.14$ & $3.04$ & $2.04$ & $0.62$ & $1.200$ & $0.667$ \\
$\rm J0533+6759$ & $5.9\times 10^{33}$ & $3.2\times 10^8$ & $4.4$ & $3.61$ & $6.4\times 10^{33}$ & $2.29$ & $1.19$ & $17.10$ & $1.72$ & $1.42$ & $0.500$ & $0.667$ \\
$\rm J0534+2200$ & $4.5\times 10^{38}$ & $5.1\times 10^{12}$ & $33.4$ & $8.45$ & $7.2\times 10^{35}$ & $0.09$ & $3.35$ & $17.80$ & $2.32$ & $1.59$ & $0.237$ & $0.610$ \\
$\rm J0540-6919$ & $1.5\times 10^{38}$ & $6.6\times 10^{12}$ & $50.6$ & $5.51$ & $7.9\times 10^{36}$ & $0.02$ & $2.24$ & $8.93$ & $2.50$ & $1.56$ & $0.348$ & $0.667$ \\
$\rm J0605+3757$ & $9.3\times 10^{33}$ & $1.5\times 10^8$ & $2.7$ & $1.58$ & $3.6\times 10^{31}$ & $1.95$ & $0.34$ & $8.78$ & $1.67$ & $1.51$ & $1.190$ & $0.667$ \\
$\rm J0610-2100$ & $8.2\times 10^{32}$ & $9.2\times 10^7$ & $3.9$ & $2.03$ & $9.2\times 10^{33}$ & $1.34$ & $0.50$ & $8.52$ & $1.81$ & $1.06$ & $0.731$ & $0.667$ \\
$\rm J0613-0200$ & $1.2\times 10^{34}$ & $2.2\times 10^8$ & $3.1$ & $3.29$ & $2.8\times 10^{33}$ & $1.16$ & $1.04$ & $11.70$ & $1.92$ & $0.97$ & $0.424$ & $0.667$ \\
$\rm J0614-3329$ & $2.2\times 10^{34}$ & $3.2\times 10^8$ & $3.1$ & $4.26$ & $5.4\times 10^{33}$ & $2.51$ & $0.04$ & $26.70$ & $1.82$ & $1.83$ & $0.536$ & $0.369$ \\
$\rm J0621+2514$ & $4.9\times 10^{34}$ & $3.5\times 10^8$ & $2.7$ & $1.57$ & $1.3\times 10^{33}$ & $2.69$ & $0.32$ & $10.30$ & $1.61$ & $2.14$ & $1.580$ & $0.667$ \\
$\rm J0631+0646$ & $1.0\times 10^{35}$ & $8.5\times 10^{11}$ & $111.0$ & $2.71$ & $4.0\times 10^{34}$ & $2.65$ & $0.77$ & $14.90$ & $1.85$ & $2.26$ & $0.907$ & $0.667$ \\
$\rm J0631+1036$ & $1.7\times 10^{35}$ & $7.4\times 10^{12}$ & $287.8$ & $3.55$ & $1.6\times 10^{34}$ & $2.00$ & $1.16$ & $16.00$ & $1.85$ & $1.54$ & $0.536$ & $0.667$ \\
$\rm J0633+0632$ & $1.2\times 10^{35}$ & $6.6\times 10^{12}$ & $297.4$ & $3.15$ & $2.1\times 10^{34}$ & $1.43$ & $0.97$ & $12.30$ & $2.04$ & $1.53$ & $0.602$ & $0.667$ \\
$\rm J0633+1746$ & $3.3\times 10^{34}$ & $2.2\times 10^{12}$ & $237.1$ & $2.83$ & $1.8\times 10^{34}$ & $1.65$ & $0.98$ & $12.10$ & $2.01$ & $1.67$ & $0.701$ & $0.696$ \\
$\rm J0659+1414$ & $3.8\times 10^{34}$ & $6.2\times 10^{12}$ & $384.9$ & $0.84$ & $2.6\times 10^{32}$ & $0.16$ & $0.13$ & $1.55$ & $2.46$ & $0.32$ & $0.799$ & $0.667$ \\
$\rm J0729-1448$ & $2.8\times 10^{35}$ & $7.2\times 10^{12}$ & $251.7$ & $4.52$ & $5.2\times 10^{33}$ & $0.07$ & $1.67$ & $7.81$ & $2.68$ & $2.02$ & $0.505$ & $0.667$ \\
$\rm J0740+6620$ & $6.0\times 10^{33}$ & $1.4\times 10^8$ & $2.9$ & $2.96$ & $4.5\times 10^{32}$ & $1.88$ & $0.89$ & $13.40$ & $1.80$ & $1.40$ & $0.604$ & $0.667$ \\
$\rm J0742-2822$ & $1.4\times 10^{35}$ & $2.3\times 10^{12}$ & $166.8$ & $3.32$ & $7.3\times 10^{33}$ & $0.95$ & $1.05$ & $10.90$ & $2.06$ & $1.08$ & $0.453$ & $0.667$ \\
$\rm J0751+1807$ & $5.7\times 10^{33}$ & $2.0\times 10^8$ & $3.5$ & $2.02$ & $1.5\times 10^{33}$ & $2.31$ & $0.50$ & $11.40$ & $1.62$ & $1.64$ & $0.983$ & $0.667$ \\
$\rm J0835-4510$ & $6.9\times 10^{36}$ & $4.5\times 10^{12}$ & $89.3$ & $4.01$ & $8.7\times 10^{34}$ & $1.19$ & $0.34$ & $15.00$ & $2.24$ & $1.91$ & $0.572$ & $0.493$ \\
$\rm J0908-4913$ & $4.9\times 10^{35}$ & $1.7\times 10^{12}$ & $106.8$ & $1.50$ & $2.6\times 10^{33}$ & $0.60$ & $0.32$ & $4.57$ & $2.64$ & $1.25$ & $1.100$ & $0.667$ \\
$\rm J0931-1902$ & $1.0\times 10^{33}$ & $1.5\times 10^8$ & $4.6$ & $1.01$ & $2.4\times 10^{33}$ & $1.89$ & $0.12$ & $5.77$ & $1.78$ & $1.75$ & $2.720$ & $0.667$ \\
$\rm J0940-5428$ & $1.9\times 10^{36}$ & $2.3\times 10^{12}$ & $87.5$ & $2.55$ & $3.0\times 10^{32}$ & $1.23$ & $0.71$ & $9.74$ & $2.08$ & $1.39$ & $0.699$ & $0.667$ \\
$\rm J0952-0607$ & $6.7\times 10^{34}$ & $1.1\times 10^8$ & $1.4$ & $1.50$ & $8.8\times 10^{33}$ & $1.83$ & $0.32$ & $8.20$ & $1.67$ & $1.42$ & $1.200$ & $0.667$ \\
$\rm J0955-6150$ & $7.1\times 10^{34}$ & $2.3\times 10^8$ & $2.0$ & $3.48$ & $4.4\times 10^{33}$ & $0.84$ & $1.13$ & $10.90$ & $2.11$ & $1.10$ & $0.438$ & $0.667$ \\
$\rm J1016-5857$ & $2.6\times 10^{36}$ & $4.0\times 10^{12}$ & $107.4$ & $5.12$ & $8.4\times 10^{34}$ & $0.79$ & $2.01$ & $15.50$ & $2.31$ & $2.02$ & $0.445$ & $0.667$ \\
$\rm J1019-5749$ & $1.9\times 10^{35}$ & $2.4\times 10^{12}$ & $162.5$ & $5.93$ & $3.9\times 10^{35}$ & $0.60$ & $2.50$ & $16.70$ & $2.37$ & $2.29$ & $0.419$ & $0.667$ \\
$\rm J1023-5746$ & $1.1\times 10^{37}$ & $8.8\times 10^{12}$ & $111.5$ & $3.48$ & $7.6\times 10^{34}$ & $0.62$ & $1.13$ & $9.73$ & $2.70$ & $2.56$ & $0.768$ & $0.667$ \\
$\rm J1024-0719^*$ & $4.6\times 10^{32}$ & $1.2\times 10^8$ & $5.2$ & $2.61$ & $7.7\times 10^{32}$ & $1.66$ & $0.73$ & $11.40$ & $1.80$ & $1.25$ & $0.634$ & $0.667$ \\
$\rm J1028-5819$ & $8.3\times 10^{35}$ & $1.6\times 10^{12}$ & $91.4$ & $4.28$ & $5.9\times 10^{34}$ & $0.97$ & $10^{-5}$ & $16.30$ & $2.37$ & $1.99$ & $0.546$ & $0.208$ \\
$\rm J1035-6720$ & $7.5\times 10^{34}$ & $4.9\times 10^8$ & $2.9$ & $2.25$ & $4.8\times 10^{33}$ & $1.65$ & $0.59$ & $10.20$ & $1.82$ & $1.32$ & $0.765$ & $0.667$ \\
$\rm J1044-5737$ & $8.0\times 10^{35}$ & $3.7\times 10^{12}$ & $139.0$ & $3.58$ & $4.9\times 10^{34}$ & $0.65$ & $1.18$ & $10.10$ & $2.24$ & $1.24$ & $0.459$ & $0.667$ \\
$\rm J1048+2339$ & $7.8\times 10^{33}$ & $4.1\times 10^8$ & $4.7$ & $4.58$ & $2.3\times 10^{33}$ & $0.89$ & $1.70$ & $14.50$ & $2.04$ & $1.02$ & $0.317$ & $0.667$ \\
$\rm J1048-5832$ & $2.0\times 10^{36}$ & $4.7\times 10^{12}$ & $123.7$ & $4.15$ & $1.9\times 10^{35}$ & $0.98$ & $0.00$ & $15.40$ & $2.32$ & $1.85$ & $0.556$ & $0.279$ \\
$\rm J1055-6028$ & $1.2\times 10^{36}$ & $2.3\times 10^{12}$ & $99.7$ & $3.97$ & $3.4\times 10^{34}$ & $1.22$ & $1.37$ & $14.20$ & $2.27$ & $2.10$ & $0.589$ & $0.667$ \\
$\rm J1057-5226$ & $3.0\times 10^{34}$ & $1.4\times 10^{12}$ & $197.1$ & $1.52$ & $3.1\times 10^{32}$ & $1.26$ & $0.91$ & $6.78$ & $2.11$ & $1.41$ & $1.040$ & $0.849$ \\
$\rm J1105-6107$ & $2.5\times 10^{36}$ & $1.4\times 10^{12}$ & $63.2$ & $3.66$ & $1.6\times 10^{34}$ & $0.52$ & $1.21$ & $9.61$ & $2.58$ & $2.08$ & $0.637$ & $0.667$ \\
$\rm J1112-6103$ & $4.5\times 10^{36}$ & $1.9\times 10^{12}$ & $65.0$ & $5.89$ & $5.3\times 10^{34}$ & $0.55$ & $2.48$ & $16.20$ & $2.69$ & $4.10$ & $0.622$ & $0.667$ \\
$\rm J1119-6127$ & $2.3\times 10^{36}$ & $5.5\times 10^{13}$ & $408.0$ & $9.67$ & $3.7\times 10^{35}$ & $0.45$ & $5.22$ & $27.70$ & $2.23$ & $2.05$ & $0.239$ & $0.667$ \\
$\rm J1124-3653^*$ & $1.7\times 10^{34}$ & $1.6\times 10^8$ & $2.4$ & $3.49$ & $1.5\times 10^{33}$ & $1.96$ & $1.13$ & $15.50$ & $1.78$ & $1.34$ & $0.497$ & $0.667$ \\
$\rm J1124-5916$ & $1.2\times 10^{37}$ & $1.4\times 10^{13}$ & $135.5$ & $3.79$ & $1.8\times 10^{35}$ & $0.36$ & $1.28$ & $8.97$ & $2.26$ & $0.95$ & $0.364$ & $0.667$ \\
$\rm J1125-5825$ & $7.9\times 10^{34}$ & $5.8\times 10^8$ & $3.1$ & $3.34$ & $2.3\times 10^{33}$ & $3.02$ & $1.06$ & $18.60$ & $1.77$ & $2.29$ & $0.743$ & $0.667$ \\
$\rm J1125-6014$ & $8.7\times 10^{33}$ & $1.4\times 10^8$ & $2.6$ & $1.77$ & $3.6\times 10^{32}$ & $1.87$ & $0.41$ & $9.22$ & $2.05$ & $1.96$ & $1.260$ & $0.667$ \\
$\rm J1137+7528$ & $8.1\times 10^{33}$ & $1.2\times 10^8$ & $2.5$ & $1.84$ & $1.1\times 10^{33}$ & $3.20$ & $0.41$ & $12.60$ & $1.62$ & $2.54$ & $1.510$ & $0.667$ \\
$\rm J1151-6108$ & $3.9\times 10^{35}$ & $1.4\times 10^{12}$ & $101.6$ & $2.49$ & $6.0\times 10^{33}$ & $1.39$ & $0.68$ & $10.10$ & $2.12$ & $1.63$ & $0.794$ & $0.667$ \\
$\rm J1207-5050$ & $1.9\times 10^{33}$ & $2.2\times 10^8$ & $4.8$ & $1.78$ & $9.2\times 10^{32}$ & $1.75$ & $0.41$ & $8.93$ & $1.85$ & $1.54$ & $1.070$ & $0.667$ \\
$\rm J1221-0633$ & $6.0\times 10^{34}$ & $2.0\times 10^8$ & $1.9$ & $3.19$ & $1.1\times 10^{33}$ & $1.48$ & $0.99$ & $12.70$ & $1.87$ & $1.16$ & $0.495$ & $0.667$ \\
$\rm J1227-4853$ & $6.9\times 10^{34}$ & $1.6\times 10^8$ & $1.7$ & $5.78$ & $5.9\times 10^{33}$ & $0.66$ & $2.41$ & $16.60$ & $2.19$ & $1.45$ & $0.317$ & $0.667$ \\
$\rm J1231-1411$ & $6.5\times 10^{33}$ & $2.3\times 10^8$ & $3.7$ & $2.63$ & $2.1\times 10^{33}$ & $1.94$ & $0.22$ & $13.20$ & $1.87$ & $1.67$ & $0.805$ & $0.526$ \\
$\rm J1253-5820$ & $5.0\times 10^{33}$ & $9.9\times 10^{11}$ & $255.5$ & $0.84$ & $1.4\times 10^{33}$ & $0.51$ & $0.13$ & $2.75$ & $2.72$ & $0.89$ & $1.570$ & $0.667$ \\
$\rm J1302-3258$ & $4.8\times 10^{33}$ & $2.1\times 10^8$ & $3.8$ & $2.38$ & $2.7\times 10^{33}$ & $2.54$ & $0.64$ & $13.40$ & $1.56$ & $1.61$ & $0.826$ & $0.667$ \\
$\rm J1311-3430$ & $4.9\times 10^{34}$ & $3.1\times 10^8$ & $2.6$ & $5.21$ & $4.3\times 10^{34}$ & $0.84$ & $0.90$ & $16.60$ & $2.15$ & $1.32$ & $0.358$ & $0.539$ \\
$\rm J1312+0051$ & $4.2\times 10^{33}$ & $2.5\times 10^8$ & $4.2$ & $2.09$ & $3.8\times 10^{33}$ & $1.46$ & $0.52$ & $9.10$ & $1.90$ & $1.30$ & $0.815$ & $0.667$ \\
$\rm J1341-6220$ & $1.4\times 10^{36}$ & $9.4\times 10^{12}$ & $193.3$ & $17.10$ & $4.2\times 10^{35}$ & - & $12.30$ & - & $2.40$ & $3.16$ & $0.180$ & $0.667$ \\
$\rm J1357-6429$ & $3.1\times 10^{36}$ & $1.0\times 10^{13}$ & $166.1$ & $1.96$ & $3.4\times 10^{34}$ & $0.35$ & $0.48$ & $4.57$ & $2.46$ & $0.88$ & $0.671$ & $0.667$ \\
$\rm J1400-1431$ & $4.3\times 10^{32}$ & $4.2\times 10^7$ & $3.1$ & $2.02$ & $5.9\times 10^{31}$ & $1.43$ & $0.50$ & $8.77$ & $1.79$ & $1.12$ & $0.761$ & $0.667$ \\
$\rm J1410-6132$ & $1.0\times 10^{37}$ & $1.7\times 10^{12}$ & $50.1$ & $2.28$ & $5.0\times 10^{35}$ & $0.75$ & $0.60$ & $7.10$ & $2.82$ & $2.22$ & $1.070$ & $0.667$ \\
$\rm J1413-6205$ & $8.3\times 10^{35}$ & $2.4\times 10^{12}$ & $109.7$ & $4.02$ & $9.8\times 10^{34}$ & $1.54$ & $0.02$ & $19.00$ & $2.28$ & $2.50$ & $0.635$ & $0.339$ \\
$\rm J1418-6058$ & $5.0\times 10^{36}$ & $5.8\times 10^{12}$ & $110.6$ & $3.97$ & $1.3\times 10^{35}$ & $1.00$ & $1.37$ & $13.10$ & $2.74$ & $3.63$ & $0.851$ & $0.667$ \\
$\rm J1420-6048$ & $1.0\times 10^{37}$ & $3.2\times 10^{12}$ & $68.2$ & $3.27$ & $4.9\times 10^{35}$ & $0.73$ & $1.03$ & $9.69$ & $2.52$ & $2.04$ & $0.704$ & $0.667$ \\
$\rm J1429-5911$ & $7.8\times 10^{35}$ & $2.5\times 10^{12}$ & $115.8$ & $3.61$ & $5.2\times 10^{34}$ & $0.54$ & $1.19$ & $9.60$ & $2.27$ & $1.19$ & $0.445$ & $0.667$ \\
$\rm J1431-4715$ & $5.5\times 10^{34}$ & $2.0\times 10^8$ & $2.0$ & $2.20$ & $1.4\times 10^{33}$ & $0.44$ & $0.57$ & $5.52$ & $2.35$ & $0.90$ & $0.605$ & $0.667$ \\
$\rm J1446-4701$ & $3.6\times 10^{34}$ & $2.0\times 10^8$ & $2.2$ & $3.93$ & $2.3\times 10^{33}$ & $1.38$ & $1.35$ & $14.80$ & $1.98$ & $1.31$ & $0.434$ & $0.667$ \\
$\rm J1455-3330$ & $1.8\times 10^{33}$ & $5.8\times 10^8$ & $8.0$ & $1.47$ & $6.4\times 10^{31}$ & $1.47$ & $0.31$ & $7.17$ & $2.41$ & $1.98$ & $1.530$ & $0.667$ \\
$\rm J1459-6053$ & $9.1\times 10^{35}$ & $2.2\times 10^{12}$ & $103.2$ & $3.80$ & $5.0\times 10^{34}$ & $0.37$ & $1.28$ & $9.03$ & $2.32$ & $1.14$ & $0.411$ & $0.667$ \\
$\rm J1509-5850$ & $5.2\times 10^{35}$ & $1.2\times 10^{12}$ & $88.9$ & $4.19$ & $1.7\times 10^{35}$ & $1.22$ & $1.49$ & $14.90$ & $2.20$ & $1.89$ & $0.520$ & $0.667$ \\
$\rm J1513-2550$ & $9.0\times 10^{34}$ & $2.9\times 10^8$ & $2.1$ & $3.20$ & $1.4\times 10^{34}$ & $1.05$ & $0.99$ & $11.00$ & $2.00$ & $1.05$ & $0.462$ & $0.667$ \\
$\rm J1514-4946$ & $9.8\times 10^{33}$ & $2.7\times 10^8$ & $3.6$ & $4.20$ & $4.1\times 10^{33}$ & $2.55$ & $1.49$ & $20.50$ & $1.74$ & $1.59$ & $0.463$ & $0.667$ \\
$\rm J1531-5610$ & $9.1\times 10^{35}$ & $1.5\times 10^{12}$ & $84.2$ & $1.41$ & $1.8\times 10^{34}$ & $1.09$ & $0.29$ & $5.89$ & $2.46$ & $1.59$ & $1.390$ & $0.667$ \\
$\rm J1536-4948$ & $2.8\times 10^{34}$ & $3.4\times 10^8$ & $3.1$ & $6.00$ & $9.1\times 10^{33}$ & $2.07$ & $1.39$ & $27.90$ & $2.04$ & $2.30$ & $0.418$ & $0.562$ \\
$\rm J1543-5149$ & $7.3\times 10^{34}$ & $2.4\times 10^8$ & $2.1$ & $3.36$ & $2.8\times 10^{33}$ & $0.64$ & $1.07$ & $9.52$ & $2.36$ & $1.48$ & $0.552$ & $0.667$ \\
$\rm J1544+4937$ & $1.2\times 10^{34}$ & $1.1\times 10^8$ & $2.2$ & $3.59$ & $2.5\times 10^{33}$ & $2.16$ & $1.18$ & $16.70$ & $1.83$ & $1.64$ & $0.553$ & $0.667$ \\
$\rm J1552+5437$ & $7.7\times 10^{33}$ & $1.1\times 10^8$ & $2.4$ & $3.45$ & $2.3\times 10^{33}$ & $1.71$ & $1.11$ & $14.50$ & $1.83$ & $1.26$ & $0.483$ & $0.667$ \\
$\rm J1555-2908$ & $3.1\times 10^{35}$ & $3.8\times 10^8$ & $1.8$ & $4.04$ & $3.2\times 10^{34}$ & $0.32$ & $1.41$ & $9.26$ & $2.52$ & $1.74$ & $0.512$ & $0.667$ \\
$\rm J1600-3053$ & $7.3\times 10^{33}$ & $2.4\times 10^8$ & $3.6$ & $3.13$ & $3.7\times 10^{33}$ & $2.74$ & $0.96$ & $16.90$ & $1.64$ & $1.72$ & $0.654$ & $0.667$ \\
$\rm J1614-2230$ & $5.0\times 10^{33}$ & $1.5\times 10^8$ & $3.2$ & $2.31$ & $1.5\times 10^{33}$ & $2.26$ & $0.61$ & $12.30$ & $1.62$ & $1.50$ & $0.811$ & $0.667$ \\
$\rm J1614-5048$ & $1.6\times 10^{36}$ & $1.4\times 10^{13}$ & $231.7$ & $3.34$ & $6.4\times 10^{34}$ & $0.97$ & $1.06$ & $11.00$ & $2.61$ & $2.67$ & $0.825$ & $0.667$ \\
$\rm J1625-0021$ & $3.7\times 10^{34}$ & $3.3\times 10^8$ & $2.8$ & $2.15$ & $2.2\times 10^{33}$ & $2.06$ & $0.55$ & $11.10$ & $1.76$ & $1.63$ & $0.917$ & $0.667$ \\
$\rm J1630+3734$ & $9.0\times 10^{33}$ & $2.2\times 10^8$ & $3.3$ & $1.90$ & $1.0\times 10^{33}$ & $1.80$ & $0.45$ & $9.46$ & $1.76$ & $1.42$ & $0.951$ & $0.667$ \\
$\rm J1640+2224$ & $1.6\times 10^{33}$ & $8.6\times 10^7$ & $3.2$ & $2.38$ & $6.2\times 10^{32}$ & $2.09$ & $0.64$ & $12.00$ & $1.84$ & $1.76$ & $0.877$ & $0.667$ \\
$\rm J1648-4611$ & $2.1\times 10^{35}$ & $2.7\times 10^{12}$ & $165.0$ & $3.89$ & $1.1\times 10^{35}$ & $2.48$ & $1.33$ & $18.90$ & $2.04$ & $2.61$ & $0.697$ & $0.667$ \\
$\rm J1653-0158$ & $4.1\times 10^{33}$ & $5.4\times 10^7$ & $2.0$ & $3.14$ & $2.9\times 10^{33}$ & $0.83$ & $0.97$ & $9.83$ & $2.00$ & $0.83$ & $0.402$ & $0.667$ \\
$\rm J1658-5324$ & $3.0\times 10^{34}$ & $2.2\times 10^8$ & $2.4$ & $1.75$ & $1.9\times 10^{33}$ & $0.82$ & $0.40$ & $5.95$ & $2.10$ & $0.93$ & $0.780$ & $0.667$ \\
$\rm J1702-4128$ & $3.4\times 10^{35}$ & $4.2\times 10^{12}$ & $182.1$ & $2.04$ & $4.9\times 10^{34}$ & $0.87$ & $0.50$ & $6.93$ & $2.84$ & $2.23$ & $1.200$ & $0.667$ \\
$\rm J1705-1906$ & $6.1\times 10^{33}$ & $1.5\times 10^{12}$ & $299.0$ & $0.50$ & $2.4\times 10^{32}$ & $0.88$ & $0.06$ & $2.79$ & $1.54$ & $0.73$ & $2.320$ & $0.667$ \\
$\rm J1709-4429$ & $3.4\times 10^{36}$ & $4.2\times 10^{12}$ & $102.5$ & $4.71$ & $1.1\times 10^{36}$ & $1.34$ & $0.20$ & $19.50$ & $2.20$ & $2.08$ & $0.509$ & $0.431$ \\
$\rm J1713+0747$ & $3.3\times 10^{33}$ & $2.6\times 10^8$ & $4.6$ & $2.54$ & $1.2\times 10^{33}$ & $2.13$ & $0.70$ & $12.70$ & $1.72$ & $1.52$ & $0.744$ & $0.667$ \\
$\rm J1718-3825$ & $1.3\times 10^{36}$ & $1.3\times 10^{12}$ & $74.7$ & $3.38$ & $1.5\times 10^{35}$ & $0.56$ & $1.08$ & $9.13$ & $2.55$ & $1.93$ & $0.654$ & $0.667$ \\
$\rm J1730-2304$ & $1.1\times 10^{33}$ & $4.6\times 10^8$ & $8.1$ & $1.20$ & $3.9\times 10^{32}$ & $0.63$ & $0.23$ & $3.95$ & $2.89$ & $1.38$ & $1.470$ & $0.667$ \\
$\rm J1730-3350$ & $1.2\times 10^{36}$ & $4.6\times 10^{12}$ & $139.5$ & $4.10$ & $6.3\times 10^{34}$ & $0.07$ & $1.44$ & $6.85$ & $2.80$ & $2.20$ & $0.590$ & $0.667$ \\
$\rm J1731-4744$ & $1.1\times 10^{34}$ & $1.6\times 10^{13}$ & $829.8$ & $0.27$ & $2.3\times 10^{32}$ & $0.37$ & $0.02$ & $1.17$ & $2.68$ & $0.47$ & $3.170$ & $0.667$ \\
$\rm J1732-3131$ & $1.5\times 10^{35}$ & $3.2\times 10^{12}$ & $196.5$ & $2.33$ & $8.8\times 10^{33}$ & $2.02$ & $0.08$ & $12.60$ & $2.24$ & $2.54$ & $1.130$ & $0.470$ \\
$\rm J1732-5049$ & $3.1\times 10^{33}$ & $3.4\times 10^8$ & $5.3$ & $2.94$ & $2.4\times 10^{33}$ & $1.34$ & $0.88$ & $11.40$ & $2.00$ & $1.33$ & $0.588$ & $0.667$ \\
$\rm J1739-3023$ & $3.0\times 10^{35}$ & $1.5\times 10^{12}$ & $114.4$ & $1.57$ & $2.4\times 10^{34}$ & $1.18$ & $0.34$ & $6.62$ & $2.62$ & $1.97$ & $1.430$ & $0.667$ \\
$\rm J1740+1000$ & $2.3\times 10^{35}$ & $2.5\times 10^{12}$ & $154.1$ & $0.69$ & $6.5\times 10^{32}$ & $0.33$ & $0.10$ & $1.87$ & $3.07$ & $0.74$ & $1.690$ & $0.667$ \\
$\rm J1741+1351$ & $2.1\times 10^{34}$ & $4.4\times 10^8$ & $3.7$ & $2.39$ & $1.2\times 10^{33}$ & $1.90$ & $0.64$ & $11.50$ & $1.75$ & $1.42$ & $0.753$ & $0.667$ \\
$\rm J1741-2054$ & $9.5\times 10^{33}$ & $3.6\times 10^{12}$ & $413.7$ & $0.98$ & $1.3\times 10^{33}$ & $0.86$ & $0.56$ & $4.32$ & $2.37$ & $1.17$ & $1.380$ & $0.858$ \\
$\rm J1744-1134$ & $4.2\times 10^{33}$ & $2.3\times 10^8$ & $4.1$ & $1.40$ & $6.9\times 10^{32}$ & $1.02$ & $0.29$ & $5.66$ & $1.99$ & $1.01$ & $1.020$ & $0.667$ \\
$\rm J1745+1017$ & $4.8\times 10^{33}$ & $1.0\times 10^8$ & $2.7$ & $2.38$ & $1.3\times 10^{33}$ & $1.80$ & $0.64$ & $11.10$ & $1.83$ & $1.47$ & $0.773$ & $0.667$ \\
$\rm J1747-2958$ & $2.5\times 10^{36}$ & $3.3\times 10^{12}$ & $98.8$ & $3.20$ & $1.2\times 10^{35}$ & $0.53$ & $0.99$ & $8.47$ & $2.93$ & $2.93$ & $0.914$ & $0.667$ \\
$\rm J1747-4036$ & $1.1\times 10^{35}$ & $2.0\times 10^8$ & $1.6$ & $2.95$ & $8.2\times 10^{34}$ & $0.82$ & $0.88$ & $9.25$ & $2.14$ & $1.10$ & $0.517$ & $0.667$ \\
$\rm J1801-2451$ & $2.6\times 10^{36}$ & $5.4\times 10^{12}$ & $124.9$ & $5.70$ & $5.5\times 10^{34}$ & $0.70$ & $2.36$ & $16.80$ & $2.32$ & $2.08$ & $0.408$ & $0.667$ \\
$\rm J1805+0615$ & $7.3\times 10^{34}$ & $2.6\times 10^8$ & $2.1$ & $2.30$ & $9.5\times 10^{33}$ & $2.05$ & $0.61$ & $11.60$ & $1.81$ & $1.67$ & $0.872$ & $0.667$ \\
$\rm J1809-2332$ & $4.3\times 10^{35}$ & $3.0\times 10^{12}$ & $146.8$ & $3.84$ & $3.9\times 10^{34}$ & $1.24$ & $0.01$ & $15.70$ & $2.30$ & $2.11$ & $0.626$ & $0.338$ \\
$\rm J1810+1744^*$ & $4.0\times 10^{34}$ & $1.2\times 10^8$ & $1.7$ & $3.62$ & $1.6\times 10^{34}$ & $0.50$ & $1.19$ & $9.41$ & $2.16$ & $0.85$ & $0.354$ & $0.667$ \\
$\rm J1811-2405$ & $2.9\times 10^{34}$ & $2.6\times 10^8$ & $2.7$ & $9.37$ & $5.5\times 10^{33}$ & $0.49$ & $4.98$ & $26.90$ & $2.34$ & $3.03$ & $0.320$ & $0.667$ \\
$\rm J1813-1246$ & $6.2\times 10^{36}$ & $1.2\times 10^{12}$ & $48.1$ & $3.32$ & $2.1\times 10^{35}$ & $0.53$ & $0.13$ & $8.23$ & $2.49$ & $1.44$ & $0.610$ & $0.452$ \\
$\rm J1816+4510$ & $5.1\times 10^{34}$ & $4.9\times 10^8$ & $3.2$ & $3.32$ & $2.4\times 10^{34}$ & $1.76$ & $1.05$ & $14.20$ & $1.78$ & $1.17$ & $0.478$ & $0.667$ \\
$\rm J1823-3021A$ & $8.3\times 10^{35}$ & $5.8\times 10^9$ & $5.4$ & $4.57$ & $1.1\times 10^{35}$ & $1.84$ & $1.70$ & $19.00$ & $1.89$ & $1.43$ & $0.397$ & $0.667$ \\
$\rm J1824+1014$ & $3.2\times 10^{33}$ & $2.0\times 10^8$ & $4.1$ & $3.88$ & $6.2\times 10^{33}$ & $2.16$ & $1.33$ & $17.80$ & $1.84$ & $1.64$ & $0.512$ & $0.667$ \\
$\rm J1824-2452A$ & $2.2\times 10^{36}$ & $3.0\times 10^9$ & $3.1$ & $3.13$ & $8.1\times 10^{34}$ & $0.95$ & $0.96$ & $10.30$ & $2.18$ & $1.35$ & $0.559$ & $0.667$ \\
$\rm J1826-1256$ & $3.6\times 10^{36}$ & $4.9\times 10^{12}$ & $110.2$ & $3.67$ & $1.2\times 10^{35}$ & $0.87$ & $1.36$ & $11.60$ & $2.46$ & $2.25$ & $0.665$ & $0.687$ \\
$\rm J1828-1101$ & $1.6\times 10^{36}$ & $1.4\times 10^{12}$ & $72.1$ & $5.54$ & $1.3\times 10^{35}$ & - & $2.27$ & - & $2.80$ & $2.21$ & $0.437$ & $0.667$ \\
$\rm J1831-0952$ & $1.1\times 10^{36}$ & $1.0\times 10^{12}$ & $67.3$ & $0.95$ & $8.8\times 10^{34}$ & $0.51$ & $0.16$ & $2.99$ & $3.15$ & $1.23$ & $1.730$ & $0.667$ \\
$\rm J1833-1034$ & $3.4\times 10^{37}$ & $4.8\times 10^{12}$ & $61.9$ & $7.70$ & $1.8\times 10^{35}$ & - & $3.71$ & - & $2.38$ & $1.51$ & $0.244$ & $0.667$ \\
$\rm J1833-3840$ & $1.1\times 10^{35}$ & $2.5\times 10^8$ & $1.9$ & $2.24$ & $7.3\times 10^{33}$ & $1.25$ & $0.58$ & $8.84$ & $2.20$ & $1.60$ & $0.871$ & $0.667$ \\
$\rm J1836+5925$ & $1.1\times 10^{34}$ & $6.9\times 10^{11}$ & $173.3$ & $2.51$ & $6.7\times 10^{33}$ & $1.47$ & $0.63$ & $10.50$ & $1.98$ & $1.43$ & $0.728$ & $0.653$ \\
$\rm J1837-0604$ & $2.0\times 10^{36}$ & $2.8\times 10^{12}$ & $96.3$ & $8.82$ & $7.5\times 10^{34}$ & $0.33$ & $4.54$ & $23.20$ & $2.63$ & $5.38$ & $0.497$ & $0.667$ \\
$\rm J1843-1113$ & $5.9\times 10^{34}$ & $1.8\times 10^8$ & $1.8$ & $3.25$ & $4.3\times 10^{33}$ & $0.21$ & $1.02$ & $6.53$ & $2.40$ & $0.94$ & $0.421$ & $0.667$ \\
$\rm J1846+0919$ & $3.4\times 10^{34}$ & $2.0\times 10^{12}$ & $225.6$ & $2.50$ & $1.0\times 10^{34}$ & $2.03$ & $0.68$ & $12.20$ & $1.73$ & $1.46$ & $0.736$ & $0.667$ \\
$\rm J1853-0004$ & $2.1\times 10^{35}$ & $1.0\times 10^{12}$ & $101.4$ & $2.56$ & $1.7\times 10^{34}$ & $3.61$ & $0.23$ & $11.20$ & $1.71$ & $3.24$ & $2.580$ & $0.667$ \\
$\rm J1855-1436$ & $9.3\times 10^{33}$ & $2.7\times 10^8$ & $3.6$ & $5.89$ & $1.6\times 10^{34}$ & $2.35$ & $2.48$ & $26.10$ & $1.96$ & $2.11$ & $0.400$ & $0.667$ \\
$\rm J1857+0143$ & $4.5\times 10^{35}$ & $2.8\times 10^{12}$ & $139.8$ & $1.40$ & $2.9\times 10^{34}$ & $2.41$ & $0.22$ & $8.45$ & $2.22$ & $2.66$ & $2.310$ & $0.667$ \\
$\rm J1858-2216$ & $1.1\times 10^{34}$ & $1.3\times 10^8$ & $2.4$ & $1.97$ & $1.2\times 10^{33}$ & $1.81$ & $0.48$ & $9.76$ & $1.77$ & $1.44$ & $0.921$ & $0.667$ \\
$\rm J1901-0125$ & $6.5\times 10^{34}$ & $4.3\times 10^8$ & $2.8$ & $3.67$ & $1.3\times 10^{34}$ & $1.14$ & $1.22$ & $12.80$ & $2.09$ & $1.39$ & $0.485$ & $0.667$ \\
$\rm J1902-5105$ & $6.6\times 10^{34}$ & $1.7\times 10^8$ & $1.7$ & $4.05$ & $7.6\times 10^{33}$ & $0.55$ & $1.41$ & $10.90$ & $2.21$ & $1.09$ & $0.374$ & $0.667$ \\
$\rm J1903-7051$ & $6.5\times 10^{33}$ & $2.3\times 10^8$ & $3.6$ & $2.52$ & $5.5\times 10^{32}$ & $2.57$ & $0.69$ & $14.00$ & $1.61$ & $1.69$ & $0.806$ & $0.667$ \\
$\rm J1907+0602$ & $2.8\times 10^{36}$ & $4.1\times 10^{12}$ & $106.6$ & $4.01$ & $2.0\times 10^{35}$ & $0.96$ & $0.00$ & $14.50$ & $2.36$ & $1.90$ & $0.579$ & $0.270$ \\
$\rm J1908+2105$ & $3.2\times 10^{34}$ & $2.5\times 10^8$ & $2.6$ & $5.21$ & $4.0\times 10^{33}$ & $1.77$ & $2.06$ & $20.90$ & $2.01$ & $1.84$ & $0.411$ & $0.667$ \\
$\rm J1913+0904$ & $1.6\times 10^{35}$ & $2.3\times 10^{12}$ & $163.2$ & $2.25$ & $2.1\times 10^{34}$ & $1.69$ & $0.59$ & $10.40$ & $2.19$ & $2.06$ & $1.030$ & $0.667$ \\
$\rm J1921+0137$ & $4.9\times 10^{34}$ & $3.0\times 10^8$ & $2.5$ & $4.38$ & $3.3\times 10^{34}$ & $1.32$ & $1.59$ & $16.00$ & $2.03$ & $1.42$ & $0.412$ & $0.667$ \\
$\rm J1921+1929$ & $7.7\times 10^{34}$ & $4.2\times 10^8$ & $2.6$ & $2.67$ & $2.5\times 10^{33}$ & $2.55$ & $0.76$ & $14.50$ & $2.03$ & $2.64$ & $1.020$ & $0.667$ \\
$\rm J1925+1720$ & $9.5\times 10^{35}$ & $1.2\times 10^{12}$ & $75.7$ & $1.87$ & $2.9\times 10^{34}$ & $1.18$ & $0.44$ & $7.50$ & $2.74$ & $2.35$ & $1.350$ & $0.667$ \\
$\rm J1932+2220$ & $7.5\times 10^{35}$ & $3.9\times 10^{12}$ & $144.5$ & $2.61$ & $7.3\times 10^{34}$ & $1.06$ & $0.73$ & $9.25$ & $2.46$ & $2.02$ & $0.874$ & $0.667$ \\
$\rm J1935+2025$ & $4.7\times 10^{36}$ & $3.0\times 10^{12}$ & $80.1$ & $1.81$ & $5.5\times 10^{34}$ & $0.55$ & $0.42$ & $5.13$ & $2.76$ & $1.51$ & $1.040$ & $0.667$ \\
$\rm J1939+2134$ & $1.1\times 10^{36}$ & $5.5\times 10^8$ & $1.6$ & $3.27$ & $3.0\times 10^{34}$ & $0.68$ & $1.02$ & $9.47$ & $2.42$ & $1.67$ & $0.616$ & $0.667$ \\
$\rm J1952+3252$ & $3.7\times 10^{36}$ & $6.5\times 10^{11}$ & $39.5$ & $3.91$ & $1.6\times 10^{35}$ & $0.94$ & $0.01$ & $13.60$ & $2.28$ & $1.62$ & $0.562$ & $0.338$ \\
$\rm J1954+2836$ & $1.1\times 10^{36}$ & $1.9\times 10^{12}$ & $92.7$ & $4.20$ & $4.9\times 10^{34}$ & $1.17$ & $1.49$ & $14.60$ & $2.11$ & $1.52$ & $0.450$ & $0.667$ \\
$\rm J1957+5033$ & $5.3\times 10^{33}$ & $2.2\times 10^{12}$ & $374.8$ & $1.17$ & $5.8\times 10^{33}$ & $0.76$ & $0.22$ & $4.26$ & $2.00$ & $0.76$ & $1.010$ & $0.667$ \\
$\rm J1958+2846$ & $3.4\times 10^{35}$ & $1.1\times 10^{13}$ & $290.4$ & $2.98$ & $4.8\times 10^{34}$ & $1.15$ & $0.89$ & $10.70$ & $2.09$ & $1.36$ & $0.588$ & $0.667$ \\
$\rm J1959+2048$ & $1.1\times 10^{35}$ & $1.9\times 10^8$ & $1.6$ & $2.08$ & $3.7\times 10^{33}$ & $1.04$ & $0.52$ & $7.66$ & $2.05$ & $1.11$ & $0.733$ & $0.667$ \\
$\rm J2006+0148$ & $1.3\times 10^{34}$ & $1.1\times 10^8$ & $2.2$ & $4.24$ & $2.2\times 10^{33}$ & $3.37$ & $1.51$ & $23.40$ & $1.68$ & $2.04$ & $0.542$ & $0.667$ \\
$\rm J2006+3102$ & $2.2\times 10^{35}$ & $2.7\times 10^{12}$ & $163.7$ & $3.13$ & $4.6\times 10^{34}$ & $1.47$ & $0.96$ & $12.50$ & $2.11$ & $1.74$ & $0.659$ & $0.667$ \\
$\rm J2017+0603$ & $1.3\times 10^{34}$ & $2.0\times 10^8$ & $2.9$ & $3.57$ & $8.3\times 10^{33}$ & $3.17$ & $1.17$ & $20.00$ & $1.59$ & $1.80$ & $0.593$ & $0.667$ \\
$\rm J2017-1614$ & $7.8\times 10^{33}$ & $1.0\times 10^8$ & $2.3$ & $4.20$ & $1.6\times 10^{33}$ & $1.60$ & $1.50$ & $16.60$ & $1.90$ & $1.28$ & $0.400$ & $0.667$ \\
$\rm J2021+3651$ & $3.4\times 10^{36}$ & $4.3\times 10^{12}$ & $103.7$ & $3.45$ & $1.9\times 10^{35}$ & $1.03$ & $0.00$ & $12.80$ & $2.36$ & $1.84$ & $0.668$ & $0.255$ \\
$\rm J2021+4026$ & $1.2\times 10^{35}$ & $5.2\times 10^{12}$ & $265.3$ & $2.82$ & $4.4\times 10^{35}$ & $0.84$ & $0.02$ & $8.82$ & $2.50$ & $1.72$ & $0.790$ & $0.386$ \\
$\rm J2022+3842$ & $3.0\times 10^{37}$ & $2.8\times 10^{12}$ & $48.6$ & $2.56$ & $3.2\times 10^{35}$ & $0.20$ & $0.71$ & $4.96$ & $2.70$ & $1.27$ & $0.652$ & $0.667$ \\
$\rm J2030+3641$ & $3.2\times 10^{34}$ & $1.5\times 10^{12}$ & $200.1$ & $2.01$ & $2.4\times 10^{35}$ & $1.87$ & $0.49$ & $10.10$ & $1.87$ & $1.65$ & $0.994$ & $0.667$ \\
$\rm J2030+4415$ & $2.2\times 10^{34}$ & $1.6\times 10^{12}$ & $227.1$ & $1.59$ & $2.7\times 10^{33}$ & $1.20$ & $0.35$ & $6.73$ & $2.07$ & $1.28$ & $1.060$ & $0.667$ \\
$\rm J2032+4127$ & $1.5\times 10^{35}$ & $1.7\times 10^{12}$ & $143.2$ & $4.15$ & $3.0\times 10^{34}$ & $2.53$ & $0.50$ & $22.50$ & $2.09$ & $2.92$ & $0.685$ & $0.521$ \\
$\rm J2039-3616$ & $9.5\times 10^{33}$ & $2.2\times 10^8$ & $3.3$ & $2.46$ & $1.3\times 10^{33}$ & $1.18$ & $0.67$ & $9.30$ & $1.91$ & $1.02$ & $0.589$ & $0.667$ \\
$\rm J2039-5617$ & $2.4\times 10^{34}$ & $2.3\times 10^8$ & $2.7$ & $4.99$ & $6.0\times 10^{33}$ & $1.47$ & $1.94$ & $18.80$ & $1.90$ & $1.10$ & $0.304$ & $0.667$ \\
$\rm J2042+0246$ & $4.7\times 10^{33}$ & $3.0\times 10^8$ & $4.5$ & $1.10$ & $2.9\times 10^{32}$ & $0.84$ & $0.20$ & $4.32$ & $2.15$ & $0.95$ & $1.250$ & $0.667$ \\
$\rm J2043+1711$ & $1.2\times 10^{34}$ & $1.3\times 10^8$ & $2.4$ & $3.95$ & $6.6\times 10^{33}$ & $1.87$ & $1.36$ & $16.90$ & $1.80$ & $1.22$ & $0.413$ & $0.667$ \\
$\rm J2043+2740$ & $5.6\times 10^{34}$ & $4.7\times 10^{11}$ & $96.1$ & $1.42$ & $2.4\times 10^{33}$ & $0.83$ & $0.29$ & $5.13$ & $2.06$ & $0.89$ & $0.927$ & $0.667$ \\
$\rm J2047+1053^*$ & $1.1\times 10^{34}$ & $4.1\times 10^8$ & $4.3$ & $6.27$ & $4.0\times 10^{33}$ & $2.75$ & $2.72$ & $29.50$ & $1.83$ & $1.77$ & $0.333$ & $0.667$ \\
$\rm J2051-0827$ & $5.2\times 10^{33}$ & $3.2\times 10^8$ & $4.5$ & $1.84$ & $6.5\times 10^{32}$ & $2.33$ & $0.43$ & $10.70$ & $1.65$ & $1.76$ & $1.130$ & $0.667$ \\
$\rm J2055+2539$ & $4.9\times 10^{33}$ & $1.5\times 10^{12}$ & $319.6$ & $1.53$ & $2.4\times 10^{33}$ & $1.21$ & $0.30$ & $6.61$ & $2.06$ & $1.28$ & $1.110$ & $0.654$ \\
$\rm J2115+5448$ & $1.7\times 10^{35}$ & $6.0\times 10^8$ & $2.6$ & $2.46$ & $8.1\times 10^{33}$ & $3.06$ & $0.67$ & $15.10$ & $1.73$ & $2.42$ & $1.050$ & $0.667$ \\
$\rm J2124-3358$ & $2.4\times 10^{33}$ & $2.6\times 10^8$ & $4.9$ & $2.02$ & $7.8\times 10^{32}$ & $1.95$ & $0.50$ & $10.30$ & $1.49$ & $1.12$ & $0.766$ & $0.667$ \\
$\rm J2208+4056$ & $8.1\times 10^{32}$ & $2.5\times 10^{12}$ & $637.0$ & $0.62$ & $3.2\times 10^{32}$ & $0.32$ & $0.09$ & $1.73$ & $2.97$ & $0.65$ & $1.720$ & $0.667$ \\
$\rm J2214+3000$ & $1.7\times 10^{34}$ & $2.7\times 10^8$ & $3.1$ & $2.22$ & $1.4\times 10^{33}$ & $1.76$ & $0.57$ & $10.50$ & $1.62$ & $1.09$ & $0.681$ & $0.667$ \\
$\rm J2215+5135$ & $7.4\times 10^{34}$ & $4.0\times 10^8$ & $2.6$ & $4.87$ & $1.7\times 10^{34}$ & $1.45$ & $1.87$ & $18.20$ & $1.94$ & $1.22$ & $0.334$ & $0.667$ \\
$\rm J2229+6114$ & $2.3\times 10^{37}$ & $2.7\times 10^{12}$ & $51.6$ & $5.12$ & $2.6\times 10^{35}$ & $0.76$ & $0.01$ & $16.70$ & $2.31$ & $1.67$ & $0.436$ & $0.308$ \\
$\rm J2234+0944$ & $4.1\times 10^{33}$ & $1.8\times 10^8$ & $3.6$ & $3.35$ & $3.0\times 10^{33}$ & $1.87$ & $1.07$ & $14.70$ & $1.80$ & $1.33$ & $0.516$ & $0.667$ \\
$\rm J2238+5903$ & $8.9\times 10^{35}$ & $5.4\times 10^{12}$ & $162.7$ & $3.72$ & $6.4\times 10^{34}$ & $0.67$ & $1.25$ & $10.70$ & $2.26$ & $1.35$ & $0.469$ & $0.667$ \\
$\rm J2240+5832$ & $2.2\times 10^{35}$ & $2.0\times 10^{12}$ & $139.9$ & $4.13$ & $6.2\times 10^{34}$ & $1.14$ & $1.46$ & $14.30$ & $2.24$ & $1.94$ & $0.537$ & $0.667$ \\
$\rm J2241-5236$ & $1.9\times 10^{34}$ & $1.4\times 10^8$ & $2.2$ & $2.28$ & $2.8\times 10^{33}$ & $1.90$ & $0.60$ & $11.10$ & $1.98$ & $1.86$ & $0.945$ & $0.667$ \\
$\rm J2256-1024$ & $3.4\times 10^{34}$ & $2.1\times 10^8$ & $2.3$ & $3.27$ & $4.2\times 10^{33}$ & $1.41$ & $1.03$ & $12.70$ & $1.84$ & $1.02$ & $0.443$ & $0.667$ \\
$\rm J2302+4442$ & $3.8\times 10^{33}$ & $3.6\times 10^8$ & $5.2$ & $3.03$ & $3.5\times 10^{33}$ & $2.48$ & $2.08$ & $14.40$ & $1.85$ & $2.05$ & $0.718$ & $0.853$ \\
$\rm J2310-0555$ & $1.1\times 10^{34}$ & $1.5\times 10^8$ & $2.6$ & $2.09$ & $1.6\times 10^{33}$ & $1.74$ & $0.52$ & $9.95$ & $1.65$ & $1.16$ & $0.753$ & $0.667$ \\
$\rm J2317+1439$ & $2.2\times 10^{33}$ & $1.2\times 10^8$ & $3.4$ & $0.55$ & $2.1\times 10^{32}$ & $1.58$ & $0.03$ & $3.52$ & $1.18$ & $1.37$ & $5.450$ & $0.667$ \\
$\rm J2339-0533$ & $2.2\times 10^{34}$ & $2.6\times 10^8$ & $2.9$ & $4.02$ & $4.2\times 10^{33}$ & $2.40$ & $1.40$ & $19.10$ & $1.61$ & $1.09$ & $0.377$ & $0.667$ \\
\enddata

\tablecomments{The parameter values have been either taken from the \citetalias{2022arXiv220111184F}, \citetalias{2005AJ....129.1993M}, and \citetalias{2013ApJS..208...17A} catalogs or derived from our meta-analysis. The 5 pulsars the $\ed$ value of which has been taken from the \citetalias{2013ApJS..208...17A} catalog are indicated by a star next to their names.}



\end{deluxetable*}


\begin{thebibliography}{}
\expandafter\ifx\csname natexlab\endcsname\relax\def\natexlab#1{#1}\fi
\providecommand{\url}[1]{\href{#1}{#1}}
\providecommand{\dodoi}[1]{doi:~\href{http://doi.org/#1}{\nolinkurl{#1}}}
\providecommand{\doeprint}[1]{\href{http://ascl.net/#1}{\nolinkurl{http://ascl.net/#1}}}
\providecommand{\doarXiv}[1]{\href{https://arxiv.org/abs/#1}{\nolinkurl{https://arxiv.org/abs/#1}}}

\bibitem[{{Abazajian}(2011)}]{Abazajian2011}
{Abazajian}, K.~N. 2011, \jcap, 2011, 010,
  \dodoi{10.1088/1475-7516/2011/03/010}

\bibitem[{{Abdo} {et~al.}(2010{\natexlab{a}}){Abdo}, {Ackermann}, {Ajello},
  {Allafort}, {Atwood}, {Baldini}, {Ballet}, {Barbiellini}, {Baring},
  {Bastieri}, {Baughman}, {Bechtol}, {Bellazzini}, {Berenji}, {Blandford},
  {Bloom}, {Bonamente}, {Borgland}, {Bouvier}, {Bregeon}, {Brez}, {Brigida},
  {Bruel}, {Burnett}, {Buson}, {Caliandro}, {Cameron}, {Caraveo}, {Carrigan},
  {Casandjian}, {Cecchi}, {{\c C}elik}, {Chekhtman}, {Cheung}, {Chiang},
  {Ciprini}, {Claus}, {Cohen-Tanugi}, {Conrad}, {Dermer}, {de Luca}, {de
  Palma}, {Dormody}, {Silva}, {Drell}, {Dubois}, {Dumora}, {Farnier},
  {Favuzzi}, {Fegan}, {Focke}, {Fortin}, {Frailis}, {Fukazawa}, {Funk},
  {Fusco}, {Gargano}, {Gasparrini}, {Gehrels}, {Germani}, {Giavitto},
  {Giebels}, {Giglietto}, {Giordano}, {Glanzman}, {Godfrey}, {Grenier},
  {Grondin}, {Grove}, {Guillemot}, {Guiriec}, {Hadasch}, {Harding}, {Hays},
  {Hobbs}, {Horan}, {Hughes}, {Jackson}, {J{\'o}hannesson}, {Johnson},
  {Johnson}, {Johnson}, {Kamae}, {Katagiri}, {Kataoka}, {Kawai}, {Kerr},
  {Kn{\"o}dlseder}, {Kuss}, {Lande}, {Latronico}, {Lee}, {Lemoine-Goumard},
  {Llena Garde}, {Longo}, {Loparco}, {Lott}, {Lovellette}, {Lubrano}, {Makeev},
  {Manchester}, {Marelli}, {Mazziotta}, {McConville}, {McEnery}, {McGlynn},
  {Meurer}, {Michelson}, {Mitthumsiri}, {Mizuno}, {Moiseev}, {Monte},
  {Monzani}, {Morselli}, {Moskalenko}, {Murgia}, {Nakamori}, {Nolan}, {Norris},
  {Noutsos}, {Nuss}, {Ohsugi}, {Omodei}, {Orlando}, {Ormes}, {Ozaki},
  {Paneque}, {Panetta}, {Parent}, {Pelassa}, {Pepe}, {Pesce-Rollins},
  {Pierbattista}, {Piron}, {Porter}, {Rain{\`o}}, {Rando}, {Ray}, {Razzano},
  {Reimer}, {Reimer}, {Reposeur}, {Ritz}, {Rochester}, {Rodriguez}, {Romani},
  {Roth}, {Ryde}, {Sadrozinski}, {Sander}, {Saz Parkinson}, {Sgr{\`o}},
  {Siskind}, {Smith}, {Smith}, {Spandre}, {Spinelli}, {Starck}, {Strickman},
  {Suson}, {Takahashi}, {Takahashi}, {Tanaka}, {Thayer}, {Thayer}, {Thompson},
  {Tibaldo}, {Torres}, {Tosti}, {Tramacere}, {Usher}, {Van Etten}, {Vasileiou},
  {Venter}, {Vilchez}, {Vitale}, {Waite}, {Wang}, {Watters}, {Weltevrede},
  {Winer}, {Wood}, {Ylinen}, \& {Ziegler}}]{2010ApJ...713..154A}
{Abdo}, A.~A., {Ackermann}, M., {Ajello}, M., {et~al.} 2010{\natexlab{a}},
  \apj, 713, 154, \dodoi{10.1088/0004-637X/713/1/154}

\bibitem[{{Abdo} {et~al.}(2010{\natexlab{b}}){Abdo}, {Ackermann}, {Ajello},
  {Baldini}, {Ballet}, {Barbiellini}, {Bastieri}, {Baughman}, {Bechtol},
  {Bellazzini}, {Berenji}, {Bignami}, {Blandford}, {Bloom}, {Bonamente},
  {Borgland}, {Bregeon}, {Brez}, {Brigida}, {Bruel}, {Burnett}, {Caliandro},
  {Cameron}, {Caraveo}, {Casandjian}, {Cecchi}, {{\c{C}}elik}, {Charles},
  {Chekhtman}, {Cheung}, {Chiang}, {Ciprini}, {Claus}, {Cohen-Tanugi},
  {Conrad}, {Dermer}, {de Palma}, {Dormody}, {Silva}, {Drell}, {Dubois},
  {Dumora}, {Edmonds}, {Farnier}, {Favuzzi}, {Fegan}, {Focke}, {Fortin},
  {Frailis}, {Fukazawa}, {Funk}, {Fusco}, {Gargano}, {Gasparrini}, {Gehrels},
  {Germani}, {Giavitto}, {Giglietto}, {Giordano}, {Glanzman}, {Godfrey},
  {Grenier}, {Grondin}, {Grove}, {Guillemot}, {Guiriec}, {Hadasch}, {Harding},
  {Hays}, {Hughes}, {J{\'o}hannesson}, {Johnson}, {Johnson}, {Johnson},
  {Kamae}, {Katagiri}, {Kataoka}, {Kawai}, {Kerr}, {Kn{\"o}dlseder}, {Kuss},
  {Lande}, {Latronico}, {Lemoine-Goumard}, {Longo}, {Loparco}, {Lott},
  {Lovellette}, {Lubrano}, {Makeev}, {Marelli}, {Mazziotta}, {McEnery},
  {Meurer}, {Michelson}, {Mitthumsiri}, {Mizuno}, {Moiseev}, {Monte},
  {Monzani}, {Morselli}, {Moskalenko}, {Murgia}, {Nolan}, {Norris}, {Nuss},
  {Ohsugi}, {Omodei}, {Orlando}, {Ormes}, {Ozaki}, {Paneque}, {Panetta},
  {Parent}, {Pelassa}, {Pepe}, {Pesce-Rollins}, {Piron}, {Porter}, {Rain{\`o}},
  {Rando}, {Ray}, {Razzano}, {Reimer}, {Reimer}, {Reposeur}, {Rochester},
  {Rodriguez}, {Romani}, {Roth}, {Ryde}, {Sadrozinski}, {Sander}, {Saz
  Parkinson}, {Scargle}, {Sgr{\`o}}, {Siskind}, {Smith}, {Smith}, {Spandre},
  {Spinelli}, {Strickman}, {Suson}, {Takahashi}, {Takahashi}, {Tanaka},
  {Thayer}, {Thayer}, {Thompson}, {Tibaldo}, {Torres}, {Tosti}, {Tramacere},
  {Usher}, {Van Etten}, {Vasileiou}, {Venter}, {Vilchez}, {Vitale}, {Waite},
  {Wang}, {Watters}, {Winer}, {Wood}, {Ylinen}, \&
  {Ziegler}}]{2010ApJ...720..272A}
---. 2010{\natexlab{b}}, \apj, 720, 272, \dodoi{10.1088/0004-637X/720/1/272}

\bibitem[{{Abdo} {et~al.}(2010{\natexlab{c}}){Abdo}, {Ajello}, {Antolini},
  {Baldini}, {Ballet}, {Barbiellini}, {Baring}, {Bastieri}, {Bechtol},
  {Bellazzini}, {Berenji}, {Bonamente}, {Borgland}, {Bouvier}, {Bregeon},
  {Brez}, {Brigida}, {Bruel}, {Buehler}, {Burnett}, {Buson}, {Caliandro},
  {Camilo}, {Caraveo}, {{\c{C}}elik}, {Chekhtman}, {Cheung}, {Chiang},
  {Ciprini}, {Claus}, {Cognard}, {Cohen-Tanugi}, {Dermer}, {de Palma}, {Digel},
  {Silva}, {Drell}, {Dubois}, {Dumora}, {Favuzzi}, {Ferrara}, {Fortin},
  {Frailis}, {Freire}, {Fukazawa}, {Funk}, {Fusco}, {Gargano}, {Gehrels},
  {Germani}, {Giglietto}, {Giordano}, {Giroletti}, {Glanzman}, {Godfrey},
  {Grenier}, {Grondin}, {Grove}, {Guillemot}, {Guiriec}, {Hadasch}, {Hanabata},
  {Harding}, {Hays}, {J{\'o}hannesson}, {Johnson}, {Johnson}, {Johnson},
  {Johnston}, {Kamae}, {Katagiri}, {Kataoka}, {Keith}, {Kerr},
  {Kn{\"o}dlseder}, {Kramer}, {Kuss}, {Lande}, {Latronico}, {Lee},
  {Lemoine-Goumard}, {Longo}, {Loparco}, {Lott}, {Lubrano}, {Makeev},
  {Manchester}, {Marelli}, {Mazziotta}, {Mitthumsiri}, {Mizuno}, {Moiseev},
  {Monte}, {Monzani}, {Morselli}, {Moskalenko}, {Murgia}, {Nakamori}, {Nolan},
  {Norris}, {Noutsos}, {Nuss}, {Ohsugi}, {Okumura}, {Orlando}, {Ormes},
  {Ozaki}, {Panetta}, {Parent}, {Pelassa}, {Pepe}, {Pesce-Rollins}, {Piron},
  {Rain{\`o}}, {Razzano}, {Reimer}, {Reimer}, {Reposeur}, {Ripken}, {Romani},
  {Sadrozinski}, {Sander}, {Saz Parkinson}, {Sgr{\`o}}, {Siskind}, {Smith},
  {Smith}, {Spandre}, {Spinelli}, {Strickman}, {Suson}, {Takahashi}, {Tanaka},
  {Thayer}, {Thayer}, {Theureau}, {Thompson}, {Thorsett}, {Tibaldo}, {Tibolla},
  {Torres}, {Tosti}, {Tramacere}, {Usher}, {Vandenbroucke}, {Vasileiou},
  {Vitale}, {Waite}, {Wang}, {Weltevrede}, {Winer}, {Yang}, {Ylinen}, \&
  {Ziegler}}]{2010ApJ...720...26A}
{Abdo}, A.~A., {Ajello}, M., {Antolini}, E., {et~al.} 2010{\natexlab{c}}, \apj,
  720, 26, \dodoi{10.1088/0004-637X/720/1/26}

\bibitem[{{Abdo} {et~al.}(2013){Abdo}, {Ajello}, {Allafort}, {Baldini},
  {Ballet}, {Barbiellini}, {Baring}, {Bastieri}, {Belfiore}, {Bellazzini}, \&
  et~al.}]{2013ApJS..208...17A}
{Abdo}, A.~A., {Ajello}, M., {Allafort}, A., {et~al.} 2013, \apjs, 208, 17,
  \dodoi{10.1088/0067-0049/208/2/17}

\bibitem[{{Abdollahi} {et~al.}(2020){Abdollahi}, {Acero}, {Ackermann},
  {Ajello}, {Atwood}, {Axelsson}, {Baldini}, {Ballet}, {Barbiellini},
  {Bastieri}, {Becerra Gonzalez}, {Bellazzini}, {Berretta}, {Bissaldi},
  {Blandford}, {Bloom}, {Bonino}, {Bottacini}, {Brandt}, {Bregeon}, {Bruel},
  {Buehler}, {Burnett}, {Buson}, {Cameron}, {Caputo}, {Caraveo}, {Casandjian},
  {Castro}, {Cavazzuti}, {Charles}, {Chaty}, {Chen}, {Cheung}, {Chiaro},
  {Ciprini}, {Cohen-Tanugi}, {Cominsky}, {Coronado-Bl{\'a}zquez}, {Costantin},
  {Cuoco}, {Cutini}, {D'Ammando}, {DeKlotz}, {de la Torre Luque}, {de Palma},
  {Desai}, {Digel}, {Di Lalla}, {Di Mauro}, {Di Venere}, {Dom{\'\i}nguez},
  {Dumora}, {Fana Dirirsa}, {Fegan}, {Ferrara}, {Franckowiak}, {Fukazawa},
  {Funk}, {Fusco}, {Gargano}, {Gasparrini}, {Giglietto}, {Giommi}, {Giordano},
  {Giroletti}, {Glanzman}, {Green}, {Grenier}, {Griffin}, {Grondin}, {Grove},
  {Guiriec}, {Harding}, {Hayashi}, {Hays}, {Hewitt}, {Horan},
  {J{\'o}hannesson}, {Johnson}, {Kamae}, {Kerr}, {Kocevski}, {Kovac'evic'},
  {Kuss}, {Landriu}, {Larsson}, {Latronico}, {Lemoine-Goumard}, {Li},
  {Liodakis}, {Longo}, {Loparco}, {Lott}, {Lovellette}, {Lubrano}, {Madejski},
  {Maldera}, {Malyshev}, {Manfreda}, {Marchesini}, {Marcotulli},
  {Mart{\'\i}-Devesa}, {Martin}, {Massaro}, {Mazziotta}, {McEnery}, {Mereu},
  {Meyer}, {Michelson}, {Mirabal}, {Mizuno}, {Monzani}, {Morselli},
  {Moskalenko}, {Negro}, {Nuss}, {Ojha}, {Omodei}, {Orienti}, {Orlando},
  {Ormes}, {Palatiello}, {Paliya}, {Paneque}, {Pei}, {Pe{\~n}a-Herazo},
  {Perkins}, {Persic}, {Pesce-Rollins}, {Petrosian}, {Petrov}, {Piron}, {Poon},
  {Porter}, {Principe}, {Rain{\`o}}, {Rando}, {Razzano}, {Razzaque}, {Reimer},
  {Reimer}, {Remy}, {Reposeur}, {Romani}, {Saz Parkinson}, {Schinzel},
  {Serini}, {Sgr{\`o}}, {Siskind}, {Smith}, {Spandre}, {Spinelli}, {Strong},
  {Suson}, {Tajima}, {Takahashi}, {Tak}, {Thayer}, {Thompson}, {Tibaldo},
  {Torres}, {Torresi}, {Valverde}, {Van Klaveren}, {van Zyl}, {Wood},
  {Yassine}, \& {Zaharijas}}]{2020ApJS..247...33A}
{Abdollahi}, S., {Acero}, F., {Ackermann}, M., {et~al.} 2020, \apjs, 247, 33,
  \dodoi{10.3847/1538-4365/ab6bcb}

\bibitem[{{Ackermann} {et~al.}(2017){Ackermann}, {Ajello}, {Albert}, {Atwood},
  {Baldini}, {Ballet}, {Barbiellini}, {Bastieri}, {Bellazzini}, {Bissaldi},
  {Blandford}, {Bloom}, {Bonino}, {Bottacini}, {Brandt}, {Bregeon}, {Bruel},
  {Buehler}, {Burnett}, {Cameron}, {Caputo}, {Caragiulo}, {Caraveo},
  {Cavazzuti}, {Cecchi}, {Charles}, {Chekhtman}, {Chiang}, {Chiappo}, {Chiaro},
  {Ciprini}, {Conrad}, {Costanza}, {Cuoco}, {Cutini}, {D'Ammando}, {de Palma},
  {Desiante}, {Digel}, {Di Lalla}, {Di Mauro}, {Di Venere}, {Drell}, {Favuzzi},
  {Fegan}, {Ferrara}, {Focke}, {Franckowiak}, {Fukazawa}, {Funk}, {Fusco},
  {Gargano}, {Gasparrini}, {Giglietto}, {Giordano}, {Giroletti}, {Glanzman},
  {Gomez-Vargas}, {Green}, {Grenier}, {Grove}, {Guillemot}, {Guiriec},
  {Gustafsson}, {Harding}, {Hays}, {Hewitt}, {Horan}, {Jogler}, {Johnson},
  {Kamae}, {Kocevski}, {Kuss}, {La Mura}, {Larsson}, {Latronico}, {Li},
  {Longo}, {Loparco}, {Lovellette}, {Lubrano}, {Magill}, {Maldera}, {Malyshev},
  {Manfreda}, {Martin}, {Mazziotta}, {Michelson}, {Mirabal}, {Mitthumsiri},
  {Mizuno}, {Moiseev}, {Monzani}, {Morselli}, {Negro}, {Nuss}, {Ohsugi},
  {Orienti}, {Orlando}, {Ormes}, {Paneque}, {Perkins}, {Persic},
  {Pesce-Rollins}, {Piron}, {Principe}, {Rain{\`o}}, {Rando}, {Razzano},
  {Razzaque}, {Reimer}, {Reimer}, {S{\'a}nchez-Conde}, {Sgr{\`o}}, {Simone},
  {Siskind}, {Spada}, {Spandre}, {Spinelli}, {Suson}, {Tajima}, {Tanaka},
  {Thayer}, {Tibaldo}, {Torres}, {Troja}, {Uchiyama}, {Vianello}, {Wood},
  {Wood}, {Zaharijas}, {Zimmer}, \& {Fermi LAT
  Collaboration}}]{2017ApJ...840...43A}
{Ackermann}, M., {Ajello}, M., {Albert}, A., {et~al.} 2017, \apj, 840, 43,
  \dodoi{10.3847/1538-4357/aa6cab}

\bibitem[{{Bai} \& {Spitkovsky}(2010)}]{2010ApJ...715.1282B}
{Bai}, X.-N., \& {Spitkovsky}, A. 2010, \apj, 715, 1282,
  \dodoi{10.1088/0004-637X/715/2/1282}

\bibitem[{{Bartels} {et~al.}(2016){Bartels}, {Krishnamurthy}, \&
  {Weniger}}]{2016PhRvL.116e1102B}
{Bartels}, R., {Krishnamurthy}, S., \& {Weniger}, C. 2016, \prl, 116, 051102,
  \dodoi{10.1103/PhysRevLett.116.051102}

\bibitem[{{Berteaud} {et~al.}(2021){Berteaud}, {Calore}, {Clavel}, {Serpico},
  {Dubus}, \& {Petrucci}}]{2021PhRvD.104d3007B}
{Berteaud}, J., {Calore}, F., {Clavel}, M., {et~al.} 2021, \prd, 104, 043007,
  \dodoi{10.1103/PhysRevD.104.043007}

\bibitem[{{Brandt} \& {Kocsis}(2015)}]{2015ApJ...812...15B}
{Brandt}, T.~D., \& {Kocsis}, B. 2015, \apj, 812, 15,
  \dodoi{10.1088/0004-637X/812/1/15}

\bibitem[{{Cerutti} {et~al.}(2016){Cerutti}, {Philippov}, \&
  {Spitkovsky}}]{2016MNRAS.457.2401C}
{Cerutti}, B., {Philippov}, A.~A., \& {Spitkovsky}, A. 2016, \mnras, 457, 2401,
  \dodoi{10.1093/mnras/stw124}

\bibitem[{{Contopoulos} \& {Kalapotharakos}(2010)}]{2010MNRAS.404..767C}
{Contopoulos}, I., \& {Kalapotharakos}, C. 2010, \mnras, 404, 767,
  \dodoi{10.1111/j.1365-2966.2010.16338.x}

\bibitem[{{DeCesar}(2013)}]{2013PhDT.......182D}
{DeCesar}, M.~E. 2013, PhD thesis, University of Maryland, College Park, United
  States

\bibitem[{{Fermi-LAT collaboration} {et~al.}(2022){Fermi-LAT collaboration},
  {:}, {Abdollahi}, {Acero}, {Baldini}, {Ballet}, {Bastieri}, {Bellazzini},
  {Berenji}, {Berretta}, {Bissaldi}, {Blandford}, {Bloom}, {Bonino}, {Brill},
  {Britto}, {Bruel}, {Burnett}, {Buson}, {Cameron}, {Caputo}, {Caraveo},
  {Castro}, {Chaty}, {Cheung}, {Chiaro}, {Cibrario}, {Ciprini},
  {Coronado-Blazquez}, {Crnogorcevic}, {Cutini}, {D'Ammando}, {De Gaetano},
  {Digel}, {Di Lalla}, {Dirirsa}, {Di Venere}, {Dominguez}, {Fallah Ramazani},
  {Fegan}, {Ferrara}, {Fiori}, {Fleischhack}, {Franckowiak}, {Fukazawa},
  {Funk}, {Fusco}, {Galanti}, {Gammaldi}, {Gargano}, {Garrappa}, {Gasparrini},
  {Giacchino}, {Giglietto}, {Giordano}, {Giroletti}, {Glanzman}, {Green},
  {Grenier}, {Grondin}, {Guillemot}, {Guiriec}, {Gustafsson}, {Harding},
  {Hays}, {Hewitt}, {Horan}, {Hou}, {Johannesson}, {Karwin}, {Kayanoki},
  {Kerr}, {Kuss}, {Landriu}, {Larsson}, {Latronico}, {Lemoine-Goumard}, {Li},
  {Liodakis}, {Longo}, {Loparco}, {Lott}, {Lubrano}, {Maldera}, {Malyshev},
  {Manfreda}, {Marti-Devesa}, {Mazziotta}, {Mereu}, {Meyer}, {Michelson},
  {Mirabal}, {Mitthumsiri}, {Mizuno}, {Moiseev}, {Monzani}, {Morselli},
  {Moskalenko}, {Negro}, {Nuss}, {Omodei}, {Orienti}, {Orlando}, {Paneque},
  {Pei}, {Perkins}, {Persic}, {Pesce-Rollins}, {Petrosian}, {Pillera}, {Poon},
  {Porter}, {Principe}, {Raino}, {Rando}, {Rani}, {Razzano}, {Razzaque},
  {Reimer}, {Reimer}, {Reposeur}, {Sanchez-Conde}, {Saz Parkinson}, {Scotton},
  {Serini}, {Sgro}, {Siskind}, {Smith}, {Spandre}, {Spinelli}, {Sueoka},
  {Suson}, {Tajima}, {Tak}, {Thayer}, {Thompson}, {Torres}, {Troja},
  {Valverde}, {Wood}, \& {Zaharijas}}]{2022arXiv220111184F}
{Fermi-LAT collaboration}, {:}, {Abdollahi}, S., {et~al.} 2022, arXiv e-prints,
  arXiv:2201.11184.
\newblock \doarXiv{2201.11184}

\bibitem[{{Gautam} {et~al.}(2021){Gautam}, {Crocker}, {Ferrario}, {Ruiter},
  {Ploeg}, {Gordon}, \& {Macias}}]{2021arXiv210600222G}
{Gautam}, A., {Crocker}, R.~M., {Ferrario}, L., {et~al.} 2021, arXiv e-prints,
  arXiv:2106.00222.
\newblock \doarXiv{2106.00222}

\bibitem[{{Gonthier} {et~al.}(2018){Gonthier}, {Harding}, {Ferrara},
  {Frederick}, {Mohr}, \& {Koh}}]{2018ApJ...863..199G}
{Gonthier}, P.~L., {Harding}, A.~K., {Ferrara}, E.~C., {et~al.} 2018, \apj,
  863, 199, \dodoi{10.3847/1538-4357/aad08d}

\bibitem[{{Harding} {et~al.}(2018){Harding}, {Kalapotharakos}, {Barnard}, \&
  {Venter}}]{2018ApJ...869L..18H}
{Harding}, A.~K., {Kalapotharakos}, C., {Barnard}, M., \& {Venter}, C. 2018,
  \apjl, 869, L18, \dodoi{10.3847/2041-8213/aaf3b2}

\bibitem[{{Harding} {et~al.}(2021){Harding}, {Venter}, \&
  {Kalapotharakos}}]{2021ApJ...923..194H}
{Harding}, A.~K., {Venter}, C., \& {Kalapotharakos}, C. 2021, \apj, 923, 194,
  \dodoi{10.3847/1538-4357/ac3084}

\bibitem[{Hooper \& Mohlabeng(2016)}]{Hooper_2016}
Hooper, D., \& Mohlabeng, G. 2016, Journal of Cosmology and Astroparticle
  Physics, 2016, 049, \dodoi{10.1088/1475-7516/2016/03/049}

\bibitem[{{Kalapotharakos} {et~al.}(2018){Kalapotharakos}, {Brambilla},
  {Timokhin}, {Harding}, \& {Kazanas}}]{2018ApJ...857...44K}
{Kalapotharakos}, C., {Brambilla}, G., {Timokhin}, A., {Harding}, A.~K., \&
  {Kazanas}, D. 2018, \apj, 857, 44, \dodoi{10.3847/1538-4357/aab550}

\bibitem[{{Kalapotharakos} \& {Contopoulos}(2009)}]{2009A&A...496..495K}
{Kalapotharakos}, C., \& {Contopoulos}, I. 2009, \aap, 496, 495,
  \dodoi{10.1051/0004-6361:200810281}

\bibitem[{{Kalapotharakos} {et~al.}(2014){Kalapotharakos}, {Harding}, \&
  {Kazanas}}]{2014ApJ...793...97K}
{Kalapotharakos}, C., {Harding}, A.~K., \& {Kazanas}, D. 2014, \apj, 793, 97,
  \dodoi{10.1088/0004-637X/793/2/97}

\bibitem[{{Kalapotharakos} {et~al.}(2017){Kalapotharakos}, {Harding},
  {Kazanas}, \& {Brambilla}}]{2017ApJ...842...80K}
{Kalapotharakos}, C., {Harding}, A.~K., {Kazanas}, D., \& {Brambilla}, G. 2017,
  \apj, 842, 80, \dodoi{10.3847/1538-4357/aa713a}

\bibitem[{{Kalapotharakos} {et~al.}(2019){Kalapotharakos}, {Harding},
  {Kazanas}, \& {Wadiasingh}}]{2019ApJ...883L...4K}
{Kalapotharakos}, C., {Harding}, A.~K., {Kazanas}, D., \& {Wadiasingh}, Z.
  2019, \apjl, 883, L4, \dodoi{10.3847/2041-8213/ab3e0a}

\bibitem[{{Lyutikov} {et~al.}(2012){Lyutikov}, {Otte}, \&
  {McCann}}]{2012ApJ...754...33L}
{Lyutikov}, M., {Otte}, N., \& {McCann}, A. 2012, \apj, 754, 33,
  \dodoi{10.1088/0004-637X/754/1/33}

\bibitem[{{Manchester} {et~al.}(2005){Manchester}, {Hobbs}, {Teoh}, \&
  {Hobbs}}]{2005AJ....129.1993M}
{Manchester}, R.~N., {Hobbs}, G.~B., {Teoh}, A., \& {Hobbs}, M. 2005, \aj, 129,
  1993, \dodoi{10.1086/428488}

\bibitem[{{P{\'e}tri}(2012)}]{2012MNRAS.424..605P}
{P{\'e}tri}, J. 2012, \mnras, 424, 605,
  \dodoi{10.1111/j.1365-2966.2012.21238.x}

\bibitem[{{Philippov} \& {Spitkovsky}(2018)}]{2018ApJ...855...94P}
{Philippov}, A.~A., \& {Spitkovsky}, A. 2018, \apj, 855, 94,
  \dodoi{10.3847/1538-4357/aaabbc}

\bibitem[{{Ploeg}(2021)}]{2021arXiv210908439P}
{Ploeg}, H. 2021, arXiv e-prints, arXiv:2109.08439.
\newblock \doarXiv{2109.08439}

\bibitem[{{Ploeg} {et~al.}(2020){Ploeg}, {Gordon}, {Crocker}, \&
  {Macias}}]{2020JCAP...12..035P}
{Ploeg}, H., {Gordon}, C., {Crocker}, R., \& {Macias}, O. 2020, \jcap, 2020,
  035, \dodoi{10.1088/1475-7516/2020/12/035}

\bibitem[{{Spitkovsky}(2006)}]{2006ApJ...648L..51S}
{Spitkovsky}, A. 2006, \apjl, 648, L51, \dodoi{10.1086/507518}

\bibitem[{{Tchekhovskoy} {et~al.}(2013){Tchekhovskoy}, {Spitkovsky}, \&
  {Li}}]{2013MNRAS.435L...1T}
{Tchekhovskoy}, A., {Spitkovsky}, A., \& {Li}, J.~G. 2013, \mnras, 435, L1,
  \dodoi{10.1093/mnrasl/slt076}

\bibitem[{{Torres}(2018)}]{2018NatAs...2..247T}
{Torres}, D.~F. 2018, Nature Astronomy, 2, 247,
  \dodoi{10.1038/s41550-018-0384-5}

\bibitem[{{Torres} {et~al.}(2019){Torres}, {Vigan{\`o}}, {Coti Zelati}, \&
  {Li}}]{2019MNRAS.489.5494T}
{Torres}, D.~F., {Vigan{\`o}}, D., {Coti Zelati}, F., \& {Li}, J. 2019, \mnras,
  489, 5494, \dodoi{10.1093/mnras/stz2403}

\bibitem[{{Yao} {et~al.}(2017){Yao}, {Manchester}, \&
  {Wang}}]{2017ApJ...835...29Y}
{Yao}, J.~M., {Manchester}, R.~N., \& {Wang}, N. 2017, \apj, 835, 29,
  \dodoi{10.3847/1538-4357/835/1/29}

\end{thebibliography}
\end{document}